\documentclass[10pt,letterpaper]{article}
\usepackage{./preamble}

\begin{document}
\vspace*{0.2in}

\begin{flushleft}
{\Large
\textbf\newline{Neuronal graphs: a graph theory primer for microscopic, functional networks of neurons recorded by Calcium imaging} 
}
\newline
\\
Carl J. Nelson\textsuperscript{1*} and
Stephen Bonner\textsuperscript{2}
\\
\bigskip
\textbf{1} School of Physics and Astronomy, University of Glasgow, Glasgow UK
\\
\textbf{2} School of Computing, Newcastle University, Newcastle, UK
\\
\bigskip

* chas.nelson@glasgow.ac.uk

\end{flushleft}

\section*{Abstract}

Connected networks are a fundamental structure of neurobiology.
Understanding these networks will help us elucidate the neural mechanisms of computation.
Mathematically speaking these networks are `graphs' --- structures containing objects that are connected.
In neuroscience, the objects could be regions of the brain, \eg{} fMRI data, or be individual neurons, \eg{} calcium imaging with fluorescence microscopy.
The formal study of graphs, graph theory, can provide neuroscientists with a large bank of algorithms for exploring networks.
Graph theory has already been applied in a variety of ways to fMRI data but, more recently, has begun to be applied at the scales of neurons, \eg{} from functional calcium imaging.
In this primer we explain the basics of graph theory and relate them to features of microscopic functional networks of neurons from calcium imaging --- neuronal graphs.
We explore recent examples of graph theory applied to calcium imaging and we highlight some areas where researchers new to the field could go awry.

\section*{Author Summary}

Carl J. Nelson is a Lord Kelvin Adam Smith Research Fellow in Data Science with a keen interest in smart microscopy, bioimage analysis and biological data exploration.
Chas' current work has brought him into the field of fast volumetric calcium imaging of zebrafish and he is currently working on several elements of this pipeline --- from faster image acquisition through to using graph theory to compare and contrast within and between samples.

Stephen Bonner is a post-doctoral researcher with a current focus on combining graphs with machine learning for real world data.
With a particular focus on incorporating the rich temporal evolution of graphs into predictive models and increasing the interpretability of such models by investigating what topological structure is being captured.

\section*{Networks of Neurons --- Neuronal Graphs}

Organised networks occur across all scales in neuroscience.
Broadly, we can categorise networks that involve neurons in two ways: macroscopic \vs{} microscopic; and functional \vs{} structural:
\begin{itemize}
  \item In fMRI recordings of brain activity, \emph{macroscopic, functional} networks are often extracted where entire brain regions (macroscopic) are related by their correlated activity or inhibition (functional).
  \item Diffusion MRI connectomics provides \emph{macroscopic, structural} networks where anatomical (structural) connections are determined between regions of the brain.
  \item Analysis of electron microscopy can be used to extract \emph{microscopic, structural} networks where neurons (microscopic) are related by their physical connections, \eg{S2020} synapses.
  \item Finally, cell-resolution calcium imaging provides \emph{microscopic, functional} networks where individual neurons are related by their correlated activity or inhibition --- we call these networks \textbf{neuronal graphs}. (Cells can also be grouped in to brain regions to create mesoscopic or macroscopic, functional networks.)
\end{itemize}

\begin{figure}
    \centering
    \begin{tikzpicture}
    \begin{scope}[thick,font=\tiny]
    \draw [{Latex[round]}-{Latex[round]}] (-2.5,0) -- (2.5,0) node [right]  {Scale};
    \draw (0,-3pt) -- (0,3pt);
    \draw (2,-3pt) -- (2,3pt)   node [above] {Regions};
    \draw (1,3pt) -- (1,-3pt)   node [below] {Groups};
    \draw (-1,-3pt) -- (-1,3pt) node [above] {Neurons};
    \draw (-2,3pt) -- (-2,-3pt) node [below] {Synapses};
    \node at (0,2) {Functional};
    \node at (0,-2) {Structural};
    \end{scope}
    \path [draw=none,fill=gray,semitransparent] (-1.5,-1.5) circle (1) node [align=center] {\eg{} EM};
    \path [draw=none,fill=gray,semitransparent] (1.5,-1.5) circle (1) node [align=center] {\eg{} dMRI};
    \path [draw=none,fill=gray,semitransparent] (1.5,1.5) circle (1) node [align=center] {\eg{} fMRI};
    \path [draw=Black,fill=NavyBlue!20] (-1.5,1.5) circle (1) node [align=center] {Neuronal\\Graphs\\\eg{} Ca\textsuperscript{2+}};
\end{tikzpicture}
    \caption{Networks of neurons occur across scales and can be functional or structural. \emph{Microscopic} networks (left), \ie{} at neuron or synapse scale, are usually recorded with calcium imaging or electron microscopy techniques. \emph{Macroscopic} networks (right),\ie{} recordings of indistinguishable groups of neurons or brain regions, are often recorded using MRI techniques. \emph{Neuronal Graphs} are microscopic (neuron-resolved), functional networks extracted from calcium imaging experiments.}\label{fig:quadrat}
\end{figure}
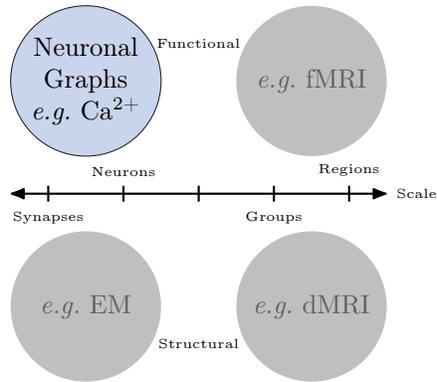

Graph theory (or network science) techniques are used for all of these network categories~\cite{BZG2018}.

This primer focusses on the application of graph theory to microscopic, functional networks of individual neurons that can be extracted from calcium imaging --- we will call these \emph{neuronal graphs}.
Analysis of neuronal graphs has shown clear differences of organisation in brain organisation through development in zebrafish~\cite{TAG2018} and xenopus~\cite{Kha2019}.
In fact, graph theory analysis was able to reveal changes in organisation of the optic tectum under dark-rearing where past experiments that used neuronal activation statistics were not~\cite{TAG2018}.
Using neuronal graphs to study development opens up greater understanding of organisational changes and, as shown by~\cite{Kha2019}, this knowledge can be used to develop, validate and compare models of specific neuronal computations.

Further, graph theory analysis allows the quantification of changes in functional organisation across the whole zebrafish brain due to genetic or pharmacological perturbations~\cite{BHGB2019}.
Combining light sheet microscopy and graph theory creates a pipeline that could be used for high-throughput small-molecule screening for the identification of new drugs.
The use of graph theory in such cases with large and densely connected neuronal graphs provides researchers with a bank of tools for exploring changes in both local and global functional organisation.

Through this paper we will briefly explain how calcium imaging data can be processed for the extraction of neuronal graphs but will focus on the essential background needed to understand and use neuronal graphs, \eg{} what types of graphs there are, \ie{} weighted \vs{} unweighted.
We will then go on to introduce some of the simpler graph theory metrics for quantifying topological structure, \eg{} degree, before moving on to some of the more complex measures available, such as centrality and community detection.
We will relate all of these theoretical notions to the underlying neuroscience and physiology being explored.
Throughout, we will highlight possible problems and challenges that the calcium imaging community should be aware of before using these graph theory techniques.
This paper will introduce common mathematical concepts from graph theory that can be applied to calcium imaging in a way that will encourage the uptake of graph theory algorithms in the field.

\section*{What is a Graph?}

\begin{table}[!t]
    \resizebox{\textwidth}{!}{
        \rowcolors{1}{white}{NavyBlue!20}
        \begin{tabular}{c p{0.5\textwidth} p{0.5\textwidth}}
            \toprule
            \textbf{Symbol} & \textbf{Definition} & \textbf{Interpretation}\\
            \midrule%
            \(G\) & A graph with an associated set of nodes \(V\) and corresponding set of edges \(E\).& A neuronal graph with a set of neurons\textsuperscript{*} and edges representing some relation between neurons.\\
            \(N_V\) & The number of nodes in the graph. & The number of neurons\textsuperscript{*} in the neuronal graph.\\
            \(N_E\) & The number of edges in the graph. & The number of edges in the neuronal graph.\\
            \(k_i\) & The total degree, or number of edges, of node \(i \in V\). & A simple measure of connectivity of a node.\\
            \(p(k)\) & The degree frequency distribution over all nodes in \(G\). & A useful summary of connectivity across a whole graph.\\
            \(A\) & The adjacency matrix, a matrix of size \(N_V \times N_V\), where \(A_{i,j}\) is \(1\) if an edge is present between nodes \(i\) and \(j\) and \(0\) otherwise. If the graph is weighted \(A_{i,j}\) is \(w\).& The inverse of the connectivity matrix (usually used in fMRI studies).\\
            \(L\) & The Laplacian matrix of graph \(G\), a matrix of size \(N_V \times N_V\). & Useful in determining metrics related to clustering and graph partitioning. \\
            \(dist(i,j)\) & The shortest path between nodes \(i\) and \(j\). & A measure of connectivity between neurons.\\
            \(\ell_G\) & The Characteristic Path Length (CPL), also known as average shortest path, of graph \(G\). & A measure of the overall information flow of a neuronal graph.\\
            \(E_G\) & The global efficiency of graph \(G\); reciprocal of \(\ell_G\). & 
            A measure of overall information flow efficiency of a neuronal graph.\\
            \(CC_i\)& The Closeness Centrality computed for node \(i\). & A measure of the importance of a neuron in neuronal graph organisation.\\
            \(B_i\)& The Betweenness Centrality computed for node \(i\). & A measure of the importance of a neuron in information flow.\\
            \(B_u\)& The Edge Betweenness Centrality computed for edge \(u\). & A measure of the importance of an edge between two neurons in information flow.\\
            \(C_i\) & The clustering coefficient for node \(i\). & A description of connectedness; less useful when considering spatial networks, \eg{} connectomics, but useful in functional networks.\\
            \(C_G\) & The global clustering coefficient for graph \(G\). & A description of the connectedness of a neuronal graph.\\
            \bottomrule
        \end{tabular}
    }
    \caption{Symbols and definitions as used in this paper. Definitions and interpretations are discussed in more detail in the main text. \textsuperscript{*} - `neurons' are often represented as regions of interest in calcium imaging data.}\label{tab:notation}
  \end{table}

\begin{figure}
    \begin{subfigure}{0.2\linewidth}
        \centering%
        \begin{tikzpicture}
    \GraphInit[vstyle=Simple]
    \SetGraphUnit{1}
    \tikzset{VertexStyle/.append style = {minimum size=5pt, fill=NavyBlue!0}}
    \Vertices{circle}{V1,V2,V3,V4,V5,V6}
    \Edge[local=true,lw=2.0pt](V1)(V2)
    \Edge[local=true,lw=2.0pt](V1)(V3)
    \Edge[local=true,lw=2.0pt](V2)(V3)
    \Edge[local=true,lw=2.0pt](V2)(V4)
    \Edge[local=true,lw=2.0pt](V2)(V5)
    \Edge[local=true,lw=2.0pt](V3)(V6)
    \Edge[local=true,lw=2.0pt](V4)(V5)
    \Edge[local=true,lw=2.0pt](V4)(V6)
\end{tikzpicture}%
        \caption{}\label{fig:graph:undirected}%
    \end{subfigure}%
    \begin{subfigure}{0.2\linewidth}
        \centering%
        \begin{tikzpicture}
    \GraphInit[vstyle=Simple]
    \SetGraphUnit{1}
    \tikzset{VertexStyle/.append style = {minimum size=5pt, fill=NavyBlue!0}}
    \Vertices{circle}{V1,V2,V3,V4,V5,V6}
    \Edge[local=true,lw=2.0pt, style={-{Latex[round]}}](V1)(V2)
    \Edge[local=true,lw=2.0pt, style={-{Latex[round]}}](V3)(V1)
    \Edge[local=true,lw=2.0pt, style={-{Latex[round]}}](V2)(V3)
    \Edge[local=true,lw=2.0pt, style={-{Latex[round]}}](V4)(V2)
    \Edge[local=true,lw=2.0pt, style={-{Latex[round]}}](V2)(V5)
    \Edge[local=true,lw=2.0pt, style={-{Latex[round]}}](V3)(V6)
    \Edge[local=true,lw=2.0pt, style={-{Latex[round]}}](V4)(V5)
    \Edge[local=true,lw=2.0pt, style={-{Latex[round]}}](V6)(V4)
\end{tikzpicture}%
        \caption{}\label{fig:graph:directed}%
    \end{subfigure}%
    \begin{subfigure}{0.2\linewidth}
        \centering%
        \begin{tikzpicture}
    \GraphInit[vstyle=Simple]
    \SetGraphUnit{1}
    \tikzset{VertexStyle/.append style = {minimum size=5pt, fill=NavyBlue!0}}
    \Vertices{circle}{V1,V2,V3,V4,V5,V6}
    \Edge[local=true,lw=2.0pt](V1)(V2)
    \Edge[local=true,lw=0.5pt](V1)(V3)
    \Edge[local=true,lw=1.0pt](V1)(V4)
    \Edge[local=true,lw=0.5pt](V1)(V5)
    \Edge[local=true,lw=1.0pt](V1)(V6)
    \Edge[local=true,lw=0.5pt](V2)(V3)
    \Edge[local=true,lw=2.0pt](V2)(V4)
    \Edge[local=true,lw=0.5pt](V2)(V5)
    \Edge[local=true,lw=1.0pt](V2)(V6)
    \Edge[local=true,lw=0.5pt](V3)(V4)
    \Edge[local=true,lw=1.0pt](V3)(V5)
    \Edge[local=true,lw=0.5pt](V3)(V6)
    \Edge[local=true,lw=2.0pt](V4)(V5)
    \Edge[local=true,lw=0.5pt](V4)(V6)
    \Edge[local=true,lw=1.0pt](V5)(V6)
\end{tikzpicture}%
        \caption{}\label{fig:graph:weight}%
    \end{subfigure}%
    \begin{subfigure}{0.375\linewidth}
        \centering%
        \input{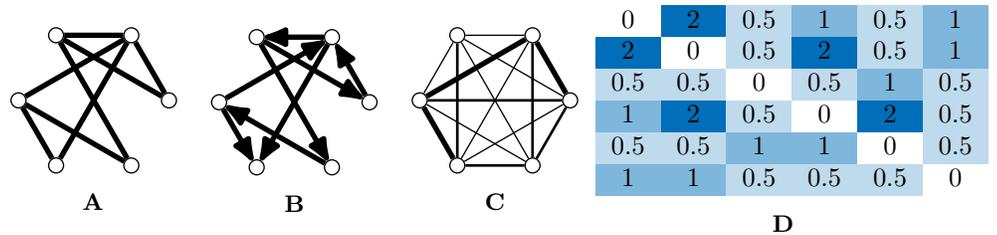}
        \caption{}\label{fig:graph:adj}%
    \end{subfigure}%
    \caption{Types of graphs and their representation.
        (\textbf{A}) Directed, unweighted graph.
        (\textbf{B}) Undirected, unweighted graph.
        (\textbf{C}) Weighted graph.
        (\textbf{D}) Adjacency matrix.
    }\label{fig:graph}%
\end{figure}

A \emph{graph} is fundamentally comprised of a set of \emph{nodes} (or vertices), with pairs of nodes connected together via an \emph{edge}.
These edges can be \emph{undirected} (\cref{fig:graph:undirected}), or \emph{directed} edges, with implied direction between two nodes creating a `directed graph' (\cref{fig:graph:directed}).
A simple graph can be defined as one which contains no self loops (edges which connect nodes with themselves) or parallel nodes (multiple edges between two nodes).

Edges (and nodes) can have associated \emph{weights}, often in the form of a numeric value.
A graph with weighted edges can be seen in \cref{fig:graph:weight}.
These graph are known as weighted graphs and are used to embed a greater quantity of information within the structure of a graph~\cite{BBPV2004}.
Working with macroscopic, functional networks, \ie{} graphs with nodes representing regions on interest,~\cite{DLFetal2020} used average edge weight as a measure of overall connectivity.

In calcium imaging a node could represent a segmented neuron and a weighted edge the strength of correlation between two nodes.
It should be noted that the term graph and network are often used interchangeably in the literature.
Mathematically a simple graph can be defined as \(G = (V,E)\) where \(V\) is a finite set of nodes or nodes and \(E\) is a set of edges~\cite{K2009}.
The elements in \(E\) are unordered pairs \( \{u,v\} \) of unique nodes \(u,v \in V\).
The number of nodes \(N_V = |V|\) and edges \(N_E = |E|\) are often called the order and size of the graph \(G\).
A directed graph \(G\) can be represented where each edge in \(E\) displays an ordering to its nodes, so that \( (u,v) \) is distinct from \( (v,u) \).

Graph theory is the study of these graphs and their mathematical properties.
Graph theory is a well developed field and provides a wide spectrum of mathematical tools for exploring and quantifying graphs.
Such graphs could be social networks, molecular modelling and, in our case, networks of neurons, \ie{} neuronal graphs, or networks of brain regions.

A graph can be represented mathematically in several forms, common ways being the adjacency, degree and Laplacian matrices~\cite{N2010}.
An adjacency matrix \(A\) for a graph \(G\), with unweighted edges, is an \(N_V\) x \(N_V\) matrix, where for a basic graph, the values are determined such that:
\begin{equation}\label{eq:adj}
    A_{ij} =
    \begin{cases}
        1 & \text{if node \(i\) and \(j\) are connected via an edge}; \\
        0 & \text{if no edge is present}.
    \end{cases}
\end{equation}

This notation can also be adjusted for the case of weighted graphs (\cref{fig:graph:weight}) such that:
\begin{equation}\label{eq:adj-weighted}
    A_{ij} =
    \begin{cases}
        w & \text{if node \(i\) and \(j\) are connected via an edge with weight \(w\)}; \\
        0 & \text{if no edge is present}.
    \end{cases}
\end{equation}

The degree \(k_i\) for any node \(i\) is the total number of edges connected to that node.
The degree distribution \(p(k)\) for graph \(G\) is the frequency of nodes with degree \(k\).

Finally, the graph Laplacian \(L\) is again a matrix of size \(N_V\) x \(N_V\). We can define the graph Laplacian matrix as,
\begin{equation}\label{eq:lapmat}
    L_{ij} = 
    \begin{cases}
        k_i & \text{if~} i = j; \\
        -1 & \text{if \(i \neq j\) and node \(i\) and \(j\) are connected by an edge};\\
        0 & \text{otherwise}.
    \end{cases}
\end{equation}

Whilst seemingly simple, the graph Laplacian has many interesting properties which can be exploited to gain insights into graph structure~\cite{C1997}.

\section*{From Calcium Imaging to Neuronal Graphs}\label{sec:extract}

The complexities and open challenges of extracting information about neurons from calcium imaging data could form a review in itself, \eg{}~\cite{PP2018}.
Here we briefly summarise the process from calcium imaging to neuronal graphs.

Calcium imaging is one of the most common ways of recording activity from large numbers of neurons at the cellular level, \cf{} electrophysiological recordings with electrodes~\cite{GK2012}.
In combination with new fluorescent reports, disease models and optogenetics, \eg{}~\cite{PRDH2015}, calcium imaging has proved a powerful tool for exploring functional activity of neurons \invitro{}, \eg{}~\cite{TVS2013} and \invivo{}, \eg{}~\cite{GLKetal2010}.
Unlike electrophysiological recordings, calcium imaging can record simultaneously from hundreds or even thousands of neurons at a time~\cite{AORetal2013}, which can lead to challenging quantities of data being produced.
However, calcium imaging does suffer from lower signal-to-noise ratio and lower temporal resolution when compared to electrophysiology which can cause issues in the extraction of neural assemblies~\cite{PP2018}.

\begin{figure}
  \begin{subfigure}[b][][b]{0.35\linewidth}
    \centering%
    \begin{tikzpicture}
    \node (image)     at (0,0) [draw, terminal] {Calcium Movie};
    \node (nuclei)    at (0,-1) [draw, predproc] {Neuron Segmentation};
    \node (correlate) at (0,-2) [draw, predproc] {Pairwise Correlation};
    \node (extract)   at (0,-3) [draw, terminal] {\textbf{Neuronal Graph}};

    \draw [-{Latex[round]}] (image) -- (nuclei);
    \draw [-{Latex[round]}] (nuclei) -- (correlate);
    \draw [-{Latex[round]}] (correlate) -- (extract);
  \end{tikzpicture}%
    \caption{}\label{fig:calcium-graphs:flow}%
  \end{subfigure}%
  \begin{subfigure}[b][][b]{0.2\linewidth}
    \centering%
    \begin{tikzpicture}
    \begin{scope}[local bounding box=graph1]
      \GraphInit[vstyle=Simple]
      \SetGraphUnit{1}
      \tikzset{VertexStyle/.append style = {minimum size = 5pt}}
      \Vertices{circle}{V1,V2,V3,V4,V5,V6}
      \Edge[local=true,lw=2.0pt](V1)(V2)
      \Edge[local=true,lw=1.5pt](V1)(V3)
      \Edge[local=true,lw=1.0pt](V1)(V4)
      \Edge[local=true,lw=0.5pt](V1)(V5)
      \Edge[local=true,lw=1.0pt](V1)(V6)
      \Edge[local=true,lw=1.5pt](V2)(V3)
      \Edge[local=true,lw=2.0pt](V2)(V4)
      \Edge[local=true,lw=1.5pt](V2)(V5)
      \Edge[local=true,lw=1.0pt](V2)(V6)
      \Edge[local=true,lw=0.5pt](V3)(V4)
      \Edge[local=true,lw=1.0pt](V3)(V5)
      \Edge[local=true,lw=1.5pt](V3)(V6)
      \Edge[local=true,lw=2.0pt](V4)(V5)
      \Edge[local=true,lw=1.5pt](V4)(V6)
      \Edge[local=true,lw=1.0pt](V5)(V6)
    \end{scope}%
    \begin{scope}[shift={(0,-2.5)},local bounding box=graph2]
      \GraphInit[vstyle=Simple]
      \SetGraphUnit{1}
      \tikzset{VertexStyle/.append style = {minimum size = 5pt}}
      \Vertices{circle}{U1,U2,U3,U4,U5,U6}
      \Edge[local=true,lw=2.0pt](U1)(U2)
      \Edge[local=true,lw=1.5pt](U1)(U3)
      \Edge[local=true,lw=1.5pt](U2)(U3)
      \Edge[local=true,lw=2.0pt](U2)(U4)
      \Edge[local=true,lw=1.5pt](U2)(U5)
      \Edge[local=true,lw=1.5pt](U3)(U6)
      \Edge[local=true,lw=2.0pt](U4)(U5)
      \Edge[local=true,lw=1.5pt](U4)(U6)
    \end{scope}%
    \draw [-{Latex[round]}, very thick] (graph1) -- (graph2) node [midway,right] {Threshold};
  \end{tikzpicture}%
    \caption{}\label{fig:calcium-graphs:threshold}%
  \end{subfigure}%
  \begin{subfigure}[b][][b]{0.4\linewidth}
    \centering%
    \begin{tikzpicture}
    \begin{scope}[local bounding box=graph1]
      \GraphInit[vstyle=Simple]
      \SetGraphUnit{1}
      \tikzset{VertexStyle/.append style = {minimum size = 5pt}}
      \Vertices{circle}{V1,V2,V3,V4,V5,V6}
      \AddVertexColor{NavyBlue}{V1}
      \AddVertexColor{NavyBlue}{V2}
      \AddVertexColor{NavyBlue}{V3}
      \Edge[local=true,lw=2.0pt,color=NavyBlue](V1)(V2)
      \Edge[local=true,lw=1.5pt,color=NavyBlue](V1)(V3)
      \Edge[local=true,lw=1.0pt](V1)(V4)
      \Edge[local=true,lw=0.5pt](V1)(V5)
      \Edge[local=true,lw=1.0pt](V1)(V6)
      \Edge[local=true,lw=1.5pt,color=NavyBlue](V2)(V3)
      \Edge[local=true,lw=2.0pt](V2)(V4)
      \Edge[local=true,lw=1.5pt](V2)(V5)
      \Edge[local=true,lw=1.0pt](V2)(V6)
      \Edge[local=true,lw=0.5pt](V3)(V4)
      \Edge[local=true,lw=1.0pt](V3)(V5)
      \Edge[local=true,lw=1.5pt](V3)(V6)
      \Edge[local=true,lw=2.0pt](V4)(V5)
      \Edge[local=true,lw=1.5pt](V4)(V6)
      \Edge[local=true,lw=1.0pt](V5)(V6)
    \end{scope}%
    \node at (2.5,0.25) {\color{NavyBlue}{Assembly \#1}};
    \begin{scope}[shift={(2.5,-1.5)},local bounding box=graph2]
      \GraphInit[vstyle=Simple]
      \SetGraphUnit{1}
      \tikzset{VertexStyle/.append style = {minimum size = 5pt}}
      \Vertices{circle}{V1,V2,V3,V4,V5,V6}
      \AddVertexColor{ForestGreen}{V2}
      \AddVertexColor{ForestGreen}{V4}
      \AddVertexColor{ForestGreen}{V5}
      \Edge[local=true,lw=2.0pt](V1)(V2)
      \Edge[local=true,lw=1.5pt](V1)(V3)
      \Edge[local=true,lw=1.0pt](V1)(V4)
      \Edge[local=true,lw=0.5pt](V1)(V5)
      \Edge[local=true,lw=1.0pt](V1)(V6)
      \Edge[local=true,lw=1.5pt](V2)(V3)
      \Edge[local=true,lw=2.0pt,color=ForestGreen](V2)(V4)
      \Edge[local=true,lw=1.5pt,color=ForestGreen](V2)(V5)
      \Edge[local=true,lw=1.0pt](V2)(V6)
      \Edge[local=true,lw=0.5pt](V3)(V4)
      \Edge[local=true,lw=1.0pt](V3)(V5)
      \Edge[local=true,lw=1.5pt](V3)(V6)
      \Edge[local=true,lw=2.0pt,color=ForestGreen](V4)(V5)
      \Edge[local=true,lw=1.5pt](V4)(V6)
      \Edge[local=true,lw=1.0pt](V5)(V6)
    \end{scope}
    \node at (0.25,-2) {\color{ForestGreen}{Assembly \#2}};
  \end{tikzpicture}%
    \caption{}\label{fig:calcium-graphs:assemblies}%
  \end{subfigure}%
  \caption{Extracting neuronal graphs from calcium imaging. (\textbf{A}) Well developed algorithms now allow for automated neuron segmentation from calcium movies. After pairwise correlation a graph is extracted. (\textbf{B}) Often neuronal graphs are thresholded, removing weak and possibly spurious edges. (\textbf{C}) Neuronal graphs can represent whole datasets, \eg{} whole-brain calcium imaging, or sub-graphs of neural assemblies, which may overlap such as in this example.}\label{fig:calcium-graphs}%
\end{figure}
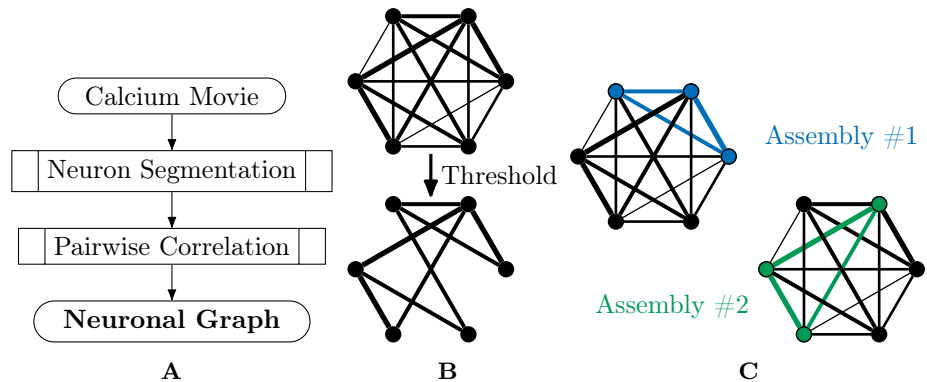

Neuronal graphs, \ie{} networks of \emph{functionally} connected neurons, can be extracted from calcium imaging with a variety of techniques, and research into accurate segmentation of neurons, processing of calcium signals and measurement of functional relation is an area of research full of caveats, warnings and open questions~\cite{SBSetal2012}.
At the simplest level, it is possible to segment individual neurons, with tools such as CaImAn~\cite{GFGetal2019}, assigning each neuron as a node, \(v_i\), in our neuronal graph.
It is then possible to measure the temporal correlation, using the Pearson correlation coefficient, of activity between pairs of neurons (\cref{fig:calcium-graphs:flow}), assigning this value as weighted edges, \(E(v_i,v_j)\), in our neuronal graph~\cite{SMU2014}.

It is important to note that a biological network of neurons is temporally directed, \ie{} the activation of one neuron causes the activation of other neurons.
However, the Pearson coefficient represents a measure of correlation insensitive to causality and, as such, neuronal graphs are usually undirected.
Very high-speed imaging of calcium dynamics (\num{>> 20} captures per second) allows the extraction of not just pairwise correlations of activity but also the propagation of calcium signalling.
By using a pairwise metric that incorporates causality, \eg{} transfer entropy, it is possible to extract directed graphs.
In~\cite{Kha2019}, the author uses such an approach to create directed neuronal graphs of small numbers of neurons in the developing Xenopus tectum to investigate looming-selective networks.

Regardless of the metric of functional connectivity used, the resulting graph will have a functional connection between with every pair of neurons.
This densely connected graph is then typically thresholded to consider only those edges (neuron-neuron correlations) with a correlation above a set value~\cite{BWAetal2019, BHGB2019} or above that expected in a random case~\cite{APMetal2017}; this removes possibly spurious neurons and connections, as well as minimising computational requirements.
Alternatively, the weakest edges are removed one-by-one until the total number of edges in the neuronal graph reaches some predetermined number~\cite{Kha2019}; this can be beneficial when comparing metrics across samples as some metrics can be skewed by the number of nodes or edges.
For whole-brain calcium imaging, \eg{} light sheet fluorescence microscopy in zebrafish, this neuronal graph represents all captured neurons (\num{\sim80000} in the zebrafish brain) and their relationship, \ie{} all overlapping neural assemblies in the brain.

However, whole-brain imaging is a relatively new technique and capturing the organisation of all neurons may not be the best way to answer a specific biological question.
In many calcium studies it makes more sense to look for sub-populations of neurons that engage in concerted activity where functionally connected neurons activate in a correlated (or anti-correlated) fashion.
These networks of functionally connected neurons, neural assemblies, may carry out specific functions, such as processing visual stimuli in the visual system; a neural assembly can be considered another type of neuronal graph.
Within a population of neurons, each neuron may be a member of multiple assemblies, ie{} multiple different neuronal graphs.
Neural assemblies have been demonstrated in a range of systems including the cortex (\eg{}~\cite{SASS2018}), hippocampus (\eg{}~\cite{HCHetal2003}) and optic tectum (\eg{}~\cite{APMetal2017}).
In particular, assemblies seen in spontaneous activity during development often demonstrate similarity to those assemblies seen in evoked stimulus processing post-development~\cite{RPPetal2015}.

Recent work summarised and compared several methods for extracting specific neural assemblies from calcium imaging data~\cite{MAG2018}.
The reader is directed to this article for further information on specific methods.
Importantly, the authors show that a spectral graph clustering approach, which does not depend directly on correlations between pairs of neurons, provided results that more consistently agreed with the consensus neural assemblies across all methods (see~\cref{box:sgc}).
Each assembly could then be considered it's own neuronal graph for further analysis.
This techniques illustrates that there may be many other future roles of graph theory in studying microscopic, functional networks of neurons.

For the purposes of this work a `neuronal graph' refers to any graph where nodes represent neurons and edges represent some measure of functional relation between pairs of neurons.

It is worth mentioning that analysing large neuronal graphs is computationally challenging and can be very noisy for further analysis.
As such, there is a large body of calcium imaging literature that groups neurons into regions of interest and using graph theory on these mesoscopic, functional networks, \eg{}~\cite{VSTetal2020, CPFBetal2019, Bet2020}.
Certain assumptions and analyses will differ between these graphs and true neuronal graphs and these assumptions will relate to how the mesoscopic networks are created, \eg{} region of interest size as discussed in~\cite{DLFetal2020}.

\begin{sidebox}[label={box:sgc}]{Similarity Graph Clustering for the Extraction of Neuronal Graphs.}
  Spectral Graph Clustering is a recently developed approach that uses powerful graph theory techniques to separate assemblies of neurons with temporally correlated activity. The technique was proposed in~\cite{APMetal2017} but, briefly, comprises the following steps:
  \begin{enumerate}
    \item Segment neurons and calculate calcium fluorescence signal (compared to baseline signal).
    \item Convert the calcium fluorescence signal to a binary activity pattern for each neuron, \ie{} at frame \(t\) neuron \(n\) is either active (\(1\)) or not (\(0\)).
    \item Identify frames where high numbers of neurons are active. Each of these frames becomes the node of a graph.
    \item Calculate the cosine-distance between the activity patterns of all pairs of frames. Edges of the above graph represent this distance metric.
    \item Use spectral clustering, a well developed graph theory method that is beyond the scope of this paper, to extract the `community structure' of this graph using statistical inference to estimate the number of communities, \ie{} assemblies.
    \item Reject certain activity patterns and communities as noise.
    \item Each neural assembly is then the average activity of all frames that belong to any kept assembly (detected community).
  \end{enumerate}
\end{sidebox}

\section*{Graph Theory Metrics}\label{sec:graphmetrics}

Once a graph has been extracted from the imaging data then a variety of metrics can be used to explore the organisation of the network.
Graph theory provides us with a range of well-defined mathematical metrics that can quantify how a graph is organised, how this evolves through time, and how the graph structure contributes to the flow of information through the network.
Changes in metrics of neuronal graphs indicate changes in the functional organisation of a system.
Such organisational changes may not be obvious when considering only population statistics of the system.
In this section we will define some of the more commonly used graph theory metrics and their relation to neuronal graphs; we will also signpost possible pitfalls for those new to interpreting graphs.

\subsection*{Node Degree}\label{sec:degree}

One of the most frequently used and easily interpretable metrics is the degree of a node.
A node's degree is simply the number of edges connected to it~\cite{N2010}.
For a directed graph, a node will have both an `in' and an `out' degree which can be calculated separately or summed together to give the total degree.
Often the degree of node \(i \in V\) is denoted by \(k_i\) and for a simple graph with \(N_v\) nodes, the degree in terms of an adjacency matrix \(A\) can be calculated as:
\begin{equation}\label{eq:degree}
    k_i = \sum_{j=1}^n A_{ij}~.
\end{equation}

\begin{figure}
    \centering
    \begin{subfigure}[b][][b]{0.275\linewidth}
        \begin{tikzpicture}
    \node at (-1.75,-0.1) [rotate=90,anchor=north] {\small{\textbf{LOW DEGREE}}};
    \begin{scope}
        \GraphInit[vstyle=Simple]
        \SetGraphUnit{1}
        \tikzset{VertexStyle/.append style = {minimum size=5pt, fill=NavyBlue!0}}
        \Vertices{circle}{V1,V2,V3,V4,V5,V6}
        \AddVertexColor{NavyBlue}{V1}
        \Edge[local=true,lw=2.0pt,color=NavyBlue](V1)(V2)
        \Edge[local=true,lw=2.0pt,color=NavyBlue](V1)(V3)
        \Edge[local=true,lw=2.0pt](V2)(V3)
        \Edge[local=true,lw=2.0pt](V2)(V4)
        \Edge[local=true,lw=2.0pt](V2)(V5)
        \Edge[local=true,lw=2.0pt](V3)(V6)
        \Edge[local=true,lw=2.0pt](V4)(V5)
        \Edge[local=true,lw=2.0pt](V4)(V6)
      \end{scope}%
      \node at (0,-1.4) {Mean: \(2.67\)};
      \node at (0,-1.8) {\color{NavyBlue}{Node: \(2\)}};
      \node at (0,-2.2) {Density: \(53 \%\)};
\end{tikzpicture}%
        \caption{}\label{fig:degree:unlow}%
    \end{subfigure}%
    \begin{subfigure}[b][][b]{0.4\linewidth}
        \centering%
        \includegraphics[width=\linewidth]{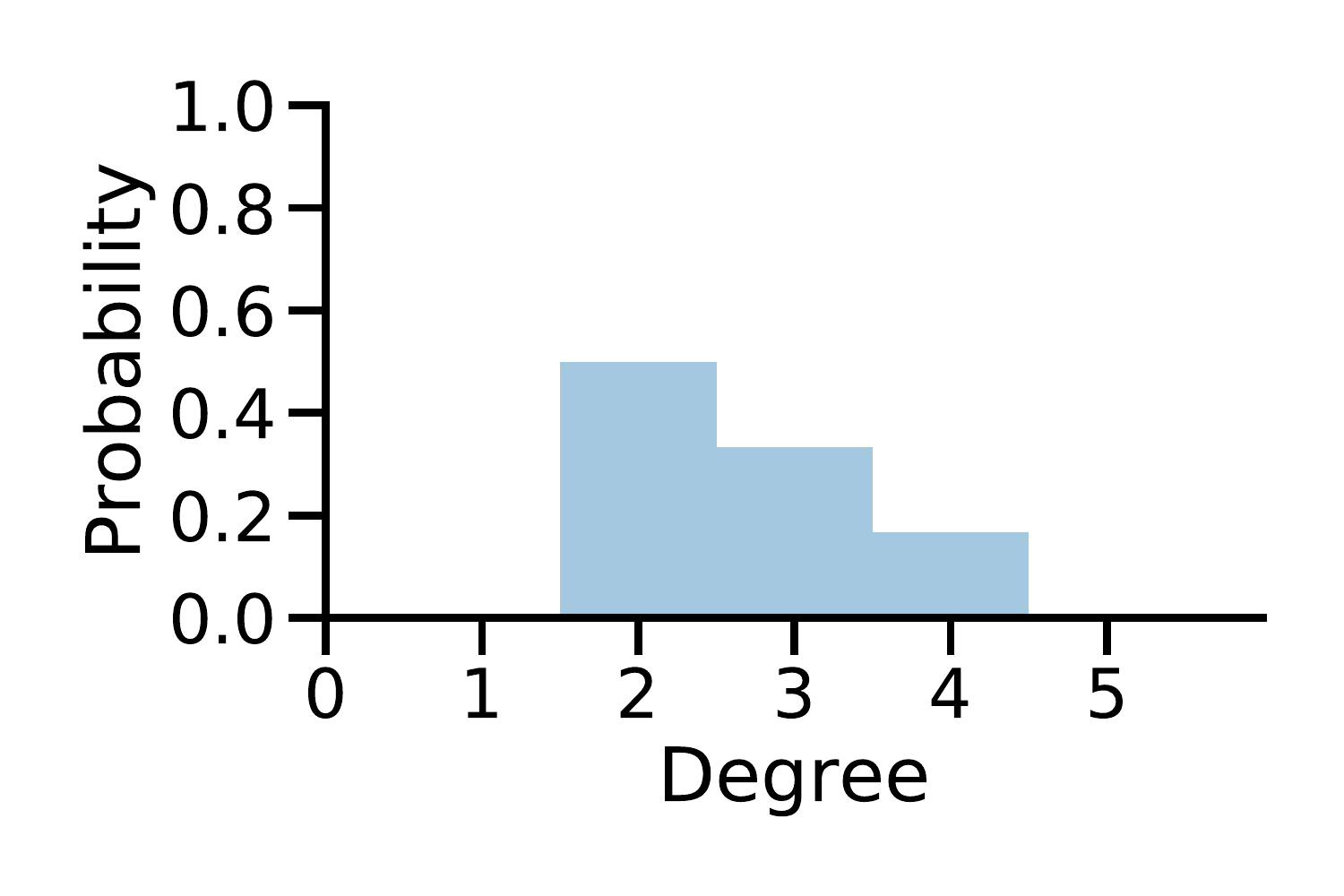}%
        \caption{}\label{fig:degree:unlow-dist}%
    \end{subfigure}\\
    \begin{subfigure}[b][][b]{0.275\linewidth}
        \begin{tikzpicture}
    \node at (-1.75,-0.1) [rotate=90,anchor=north] {\small{\textbf{HIGH DEGREE}}};
    \begin{scope}
        \GraphInit[vstyle=Simple]
        \SetGraphUnit{1}
        \tikzset{VertexStyle/.append style = {minimum size=5pt, fill=NavyBlue!0}}
        \Vertices{circle}{V1,V2,V3,V4,V5,V6}
        \AddVertexColor{NavyBlue}{V1}
        \Edge[local=true,lw=2.0pt,color=NavyBlue](V1)(V2)
        \Edge[local=true,lw=2.0pt,color=NavyBlue](V1)(V3)
        \Edge[local=true,lw=2.0pt,color=NavyBlue](V1)(V4)
        \Edge[local=true,lw=2.0pt,color=NavyBlue](V1)(V5)
        \Edge[local=true,lw=2.0pt,color=NavyBlue](V1)(V6)
        \Edge[local=true,lw=2.0pt](V2)(V3)
        \Edge[local=true,lw=2.0pt](V2)(V4)
        \Edge[local=true,lw=2.0pt](V2)(V5)
        \Edge[local=true,lw=2.0pt](V2)(V6)
        \Edge[local=true,lw=2.0pt](V3)(V4)
        \Edge[local=true,lw=2.0pt](V3)(V5)
        \Edge[local=true,lw=2.0pt](V3)(V6)
        \Edge[local=true,lw=2.0pt](V4)(V5)
        \Edge[local=true,lw=2.0pt](V4)(V6)
        \Edge[local=true,lw=2.0pt](V5)(V6)
      \end{scope}%
      \node at (0,-1.4) {Mean: \(5\)};
      \node at (0,-1.8) {\color{NavyBlue}{Node: \(5\)}};
      \node at (0,-2.2) {Density: \(100 \%\)};
\end{tikzpicture}%
        \caption{}\label{fig:degree:unhigh}%
    \end{subfigure}%
    \begin{subfigure}[b][][b]{0.4\linewidth}
        \centering%
        \includegraphics[width=\linewidth]{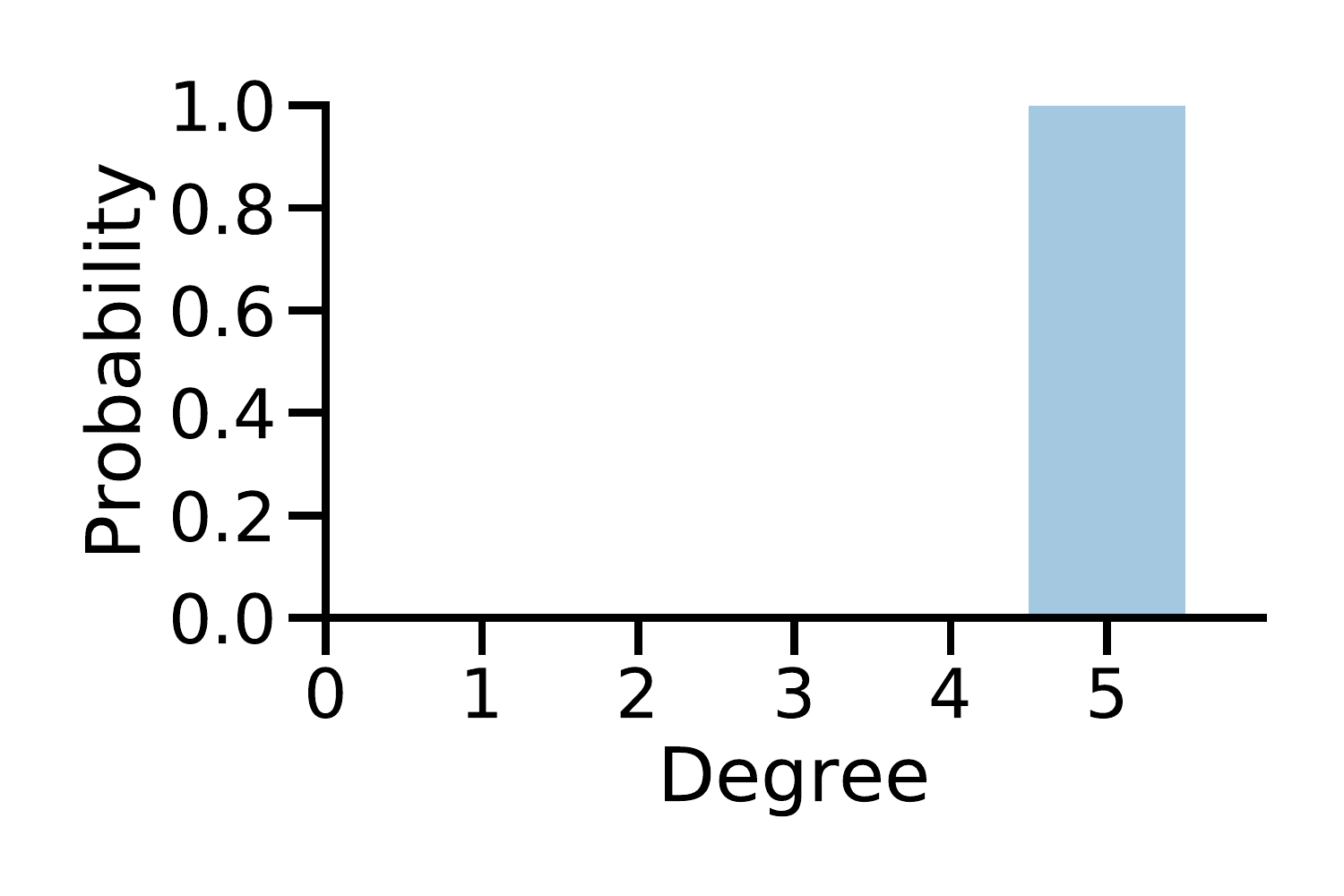}%
        \caption{}\label{fig:degree:unhigh-dist}%
    \end{subfigure}%
    \caption{Graph degree, degree distribution and density reveal information about how connected nodes in a graph are.
        (\textbf{A--B}) A graph (\textbf{A}) with low density and thus low mean degree and individual node degree as shown by the degree distribution (\textbf{B}).
        (\textbf{C--D}) A `complete' graph (\textbf{C}) with \(100\% \) density and thus high mean and node degree and a different degree distribution (\textbf{D}, \cf{} \textbf{B}).}\label{fig:degree}
\end{figure}
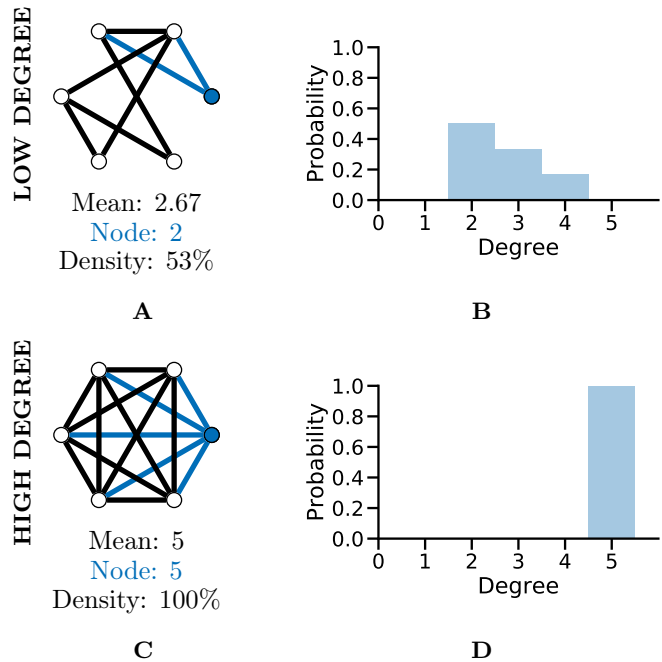

Although a simplistic metric, the graph degree alone can provide significant information about a graph.
As an example of this, comparing the two unweighted graphs in~\cref{fig:degree:unlow} and~\cref{fig:degree:unhigh}, highlights that although both graphs have the same number of nodes, the mean degree of each graph is significantly different, with the higher mean degree of~\labelcref{fig:degree:unhigh} indicating that this graph is much more densely connected than~\labelcref{fig:degree:unhigh}.

To further analyse the structure of complex graphs, the distribution of degree values is frequently used (See (\cref{fig:degree:unlow-dist} and~\cref{fig:degree:unhigh-dist})).
The degree distribution is used to calculate the probability that a randomly selected node will have a certain degree value.
It provides a natural overview of connectivity within a graph and is often plotted as a histogram with a bin size of one~\cite{N2003}. In~\cref{fig:degree:unlow-dist} for example, studying the degree distribution alone would inform us that the associated graph is fully connected, known as a complete graph. 

Another metric producing a single score indicating how connected nodes are within a graph is that of density. Graph density measures the number of existing edges within a graph versus the total number of possible edges,~\eg{}~\cref{fig:degree:unlow} and~\cref{fig:degree:unhigh}, where the density score of 100\% informs us the graph has all nodes connected to all other nodes. The interpretation of degree distribution is also important, particularly in relation to scale-free networks (see Section below).

The degree of a graph has been used as a measure of the number of functional connections between neurons and used to quantify network properties in cell cultures~\cite{SMU2014} and \invivo{}~\cite{APMetal2017,TAG2018,BHGB2019,Kha2019}.
It's common to use the mean degree of a neuronal graph, which represents a measure of overall connectivity for a system~\cite{APMetal2017}.
Throughout the development of the zebrafish optic tectum the degree of the representing neuronal graphs increases during development to a mid-development peak followed  by a slight decreased towards the end of development indicating that neuronal systems go through different phases of reorganisation during development~\cite{APMetal2017}.

\subsection*{Paths in Graphs}\label{sec:paths}

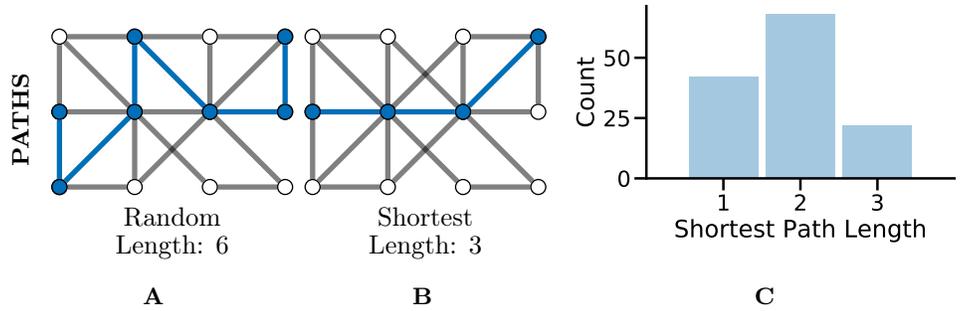
\begin{figure}
    \centering%
    \begin{subfigure}[b][][b]{0.30\linewidth}
        \begin{tikzpicture}
    \node at (-1.75,-0.1) [rotate=90,anchor=north] {\small{\textbf{PATHS}}};
    \begin{scope}[shift={(-1,0)}]
        \GraphInit[vstyle=Simple]
        \SetGraphUnit{1}
        \tikzset{VertexStyle/.append style = {minimum size=5pt, fill=NavyBlue!0}}
        N==E
        S==W
        \Vertex{V1} \NO(V1){V2} \SO(V1){V3}
        \EA(V1){V4} \NOEA(V1){V5} \SOEA(V1){V6}
        \EA(V4){V7} \NOEA(V4){V8} \SOEA(V4){V9}
        \EA(V7){V10} \NOEA(V7){V11} \SOEA(V7){V12}
        \AddVertexColor{NavyBlue}{V1}
        \AddVertexColor{NavyBlue}{V3}
        \AddVertexColor{NavyBlue}{V4}
        \AddVertexColor{NavyBlue}{V5}
        \AddVertexColor{NavyBlue}{V7}
        \AddVertexColor{NavyBlue}{V10}
        \AddVertexColor{NavyBlue}{V11}

        \Edge[local=true,lw=2.0pt,style={opacity=0.5}](V1)(V2)
        \Edge[local=true,lw=2.0pt,style={opacity=0.5}](V1)(V4)
        \Edge[local=true,lw=2.0pt,style={opacity=0.5}](V2)(V4)
        \Edge[local=true,lw=2.0pt,style={opacity=0.5}](V2)(V5)
        \Edge[local=true,lw=2.0pt,style={opacity=0.5}](V3)(V6)
        \Edge[local=true,lw=2.0pt,style={opacity=0.5}](V4)(V6)
        \Edge[local=true,lw=2.0pt,style={opacity=0.5}](V4)(V7)
        \Edge[local=true,lw=2.0pt,style={opacity=0.5}](V4)(V9)
        \Edge[local=true,lw=2.0pt,style={opacity=0.5}](V5)(V8)
        \Edge[local=true,lw=2.0pt,style={opacity=0.5}](V6)(V7)
        \Edge[local=true,lw=2.0pt,style={opacity=0.5}](V7)(V8)
        \Edge[local=true,lw=2.0pt,style={opacity=0.5}](V7)(V11)
        \Edge[local=true,lw=2.0pt,style={opacity=0.5}](V7)(V12)
        \Edge[local=true,lw=2.0pt,style={opacity=0.5}](V8)(V11)
        \Edge[local=true,lw=2.0pt,style={opacity=0.5}](V9)(V12)

        \Edge[local=true,lw=2.0pt,color=NavyBlue](V1)(V3)
        \Edge[local=true,lw=2.0pt,color=NavyBlue](V3)(V4)
        \Edge[local=true,lw=2.0pt,color=NavyBlue](V4)(V5)
        \Edge[local=true,lw=2.0pt,color=NavyBlue](V5)(V7)
        \Edge[local=true,lw=2.0pt,color=NavyBlue](V7)(V10)
        \Edge[local=true,lw=2.0pt,color=NavyBlue](V10)(V11)
      \end{scope}%
      \node at (0.5,-1.4) {Random};
      \node at (0.5,-1.8) {Length: \(6\)};
\end{tikzpicture}%
        \caption{}\label{fig:path:long}%
    \end{subfigure}%
    \begin{subfigure}[b][][b]{0.235\linewidth}
        \begin{tikzpicture}
    \begin{scope}[shift={(-1,0)}]
        \GraphInit[vstyle=Simple]
        \SetGraphUnit{1}
        \tikzset{VertexStyle/.append style = {minimum size=5pt, fill=NavyBlue!0}}
        \Vertex{V1} \NO(V1){V2} \SO(V1){V3}
        \EA(V1){V4} \NOEA(V1){V5} \SOEA(V1){V6}
        \EA(V4){V7} \NOEA(V4){V8} \SOEA(V4){V9}
        \EA(V7){V10} \NOEA(V7){V11} \SOEA(V7){V12}
        \AddVertexColor{NavyBlue}{V1}
        \AddVertexColor{NavyBlue}{V4}
        \AddVertexColor{NavyBlue}{V7}
        \AddVertexColor{NavyBlue}{V11}

        \Edge[local=true,lw=2.0pt,style={opacity=0.5}](V1)(V2)
        \Edge[local=true,lw=2.0pt,style={opacity=0.5}](V1)(V3)
        \Edge[local=true,lw=2.0pt,style={opacity=0.5}](V2)(V4)
        \Edge[local=true,lw=2.0pt,style={opacity=0.5}](V2)(V5)
        \Edge[local=true,lw=2.0pt,style={opacity=0.5}](V3)(V4)
        \Edge[local=true,lw=2.0pt,style={opacity=0.5}](V3)(V6)
        \Edge[local=true,lw=2.0pt,style={opacity=0.5}](V4)(V5)
        \Edge[local=true,lw=2.0pt,style={opacity=0.5}](V4)(V6)
        \Edge[local=true,lw=2.0pt,style={opacity=0.5}](V4)(V8)
        \Edge[local=true,lw=2.0pt,style={opacity=0.5}](V4)(V9)
        \Edge[local=true,lw=2.0pt,style={opacity=0.5}](V5)(V7)
        \Edge[local=true,lw=2.0pt,style={opacity=0.5}](V6)(V7)
        \Edge[local=true,lw=2.0pt,style={opacity=0.5}](V7)(V8)
        \Edge[local=true,lw=2.0pt,style={opacity=0.5}](V7)(V10)
        \Edge[local=true,lw=2.0pt,style={opacity=0.5}](V7)(V12)
        \Edge[local=true,lw=2.0pt,style={opacity=0.5}](V8)(V11)
        \Edge[local=true,lw=2.0pt,style={opacity=0.5}](V9)(V12)
        \Edge[local=true,lw=2.0pt,style={opacity=0.5}](V10)(V11)

        \Edge[local=true,lw=2.0pt,color=NavyBlue](V1)(V4)
        \Edge[local=true,lw=2.0pt,color=NavyBlue](V4)(V7)
        \Edge[local=true,lw=2.0pt,color=NavyBlue](V7)(V11)
      \end{scope}%
      \node at (0.5,-1.4) {Shortest};
      \node at (0.5,-1.8) {Length: \(3\)};
\end{tikzpicture}%
        \caption{}\label{fig:path:short}%
    \end{subfigure}
    \begin{subfigure}[b][][b]{0.43\linewidth}
        \centering%
        \includegraphics[width=\linewidth]{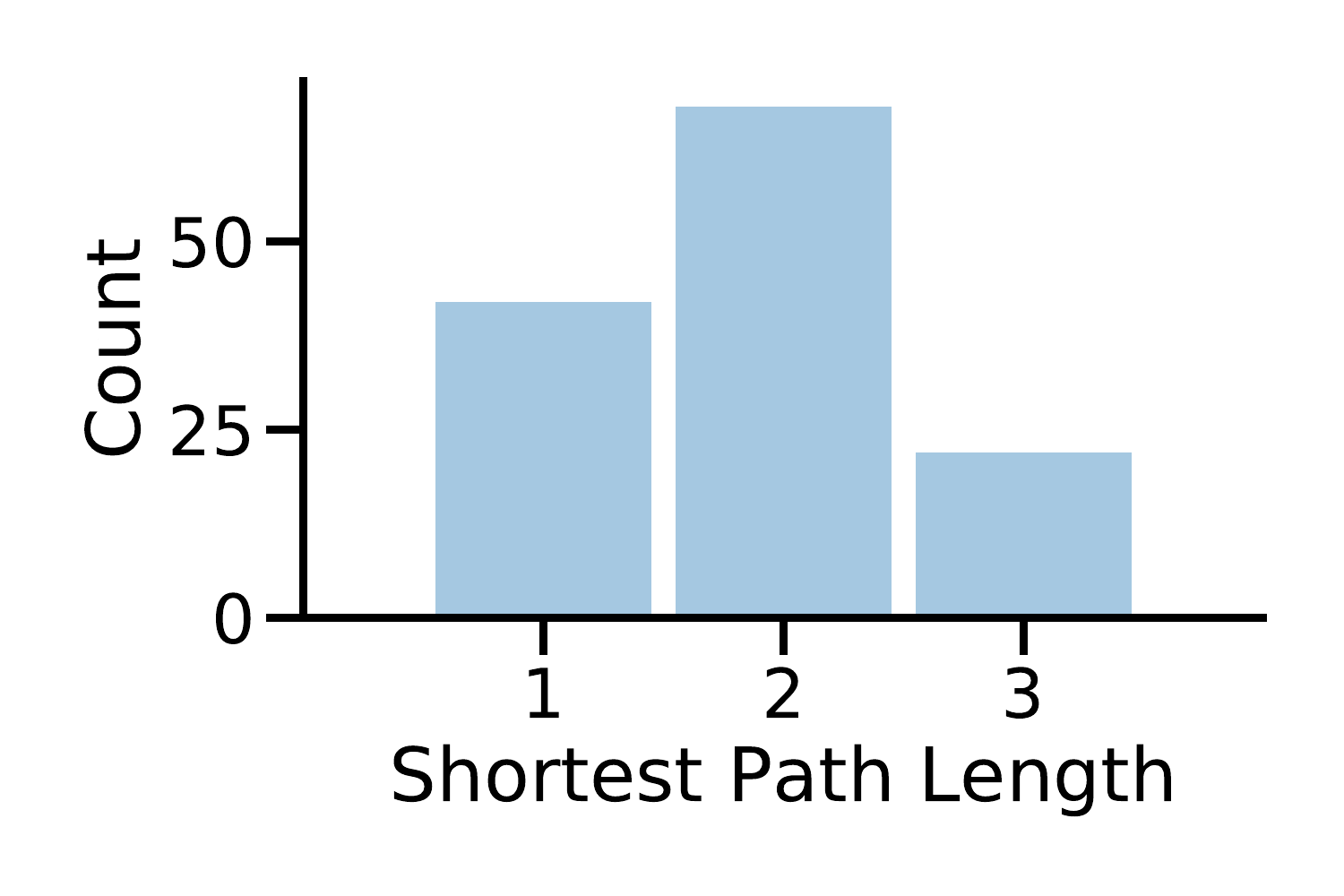}%
        \caption{}\label{fig:path:hist}%
    \end{subfigure}%
    \caption{Paths, and especially shortest paths, in graphs give an idea of efficiency of flow or information transfer.
        (\textbf{A}) An example random path through a graph between two nodes.
        (\textbf{B}) An example shortest path, there are multiple routes of the same length, between the same two nodes in the same graph.
        (\textbf{C}) The distribution of shortest path lengths across all pairs of nodes in a graph can give an idea of flow efficiency in a network.}\label{fig:path}%
\end{figure}

Another common set of graph metrics to consider revolve around the concept of a path in a graph.
A path is a route from one node to another through the graph, in such a way that every pair of nodes along the path are adjacent to one another.
A path which contains no repeated vertices is known as a simple path.
A graph for which there exists a path between every pair of nodes is considered a connected graph~\cite{K2009}, which can be seen clearly in~\cref{fig:degree:unhigh}.
Often there are many possible paths between two nodes, in which case the shortest possible path, which is the minimum number of edges needing to be traversed to connect two nodes, is often an interesting metric to consider~\cite{HJ2012}.
This concept is highlighted in~\cref{fig:path:long} and~\cref{fig:path:short} which both illustrate paths between the same two nodes within the graph, where the first is a random path and the second is the shortest possible path.

In~\cite{APMetal2017}, the authors relate the length of a path to the potential for functional integration between two nodes.
The shorter the path, the greater the potential for functional integration, \ie{} a shorter average path length implies that information can be more efficiently shared across the whole network.
In turn, the potential for functional integration is closely linked with efficiency communication between nodes, \ie{} shorter paths between nodes indicate a smaller number of functional pathways between neurons and thus more efficient communication between neurons. Although it should be noted that this being universally true has been disputed in the literature, with evidence that some information taking longer paths to retrain the correct information modality \cite{FB2015} . 

\subsubsection*{Characteristic Path Length}\label{sec:cpl}

Linked to the shortest path is the characteristic path length (CPL) or average shortest path length of a graph, as used in~\cite{BHGB2019}.
The CPL is calculated by first computing the average shortest distance for all nodes to all other nodes, then taking the mean of the resulting values:
\begin{equation}\label{eq:cpl}
\ell_G = \frac{1}{N_V(N_V-1)} \sum\limits_{ i,j \in V i \ne j} dist(i,j)~,
\end{equation}
where \(dist\) is the shortest path between node \(i\) and \(j\).
The CPL represents a measure of functional integration in a neuronal graph: a lower CPL represents short functional paths throughout the network and thus improved potential for integration and parallel processing cross the graph.

\subsubsection*{Global Graph Efficiency}\label{sec:efficiency}

In \cite{APMetal2017}, the authors use a different but related metric --- global graph efficiency.
Global graph efficiency again draws on the shortest path concept, and allows for a measure on how efficiently information can flow within a entire graph~\cite{LM2003,EVN2015}.
It can also be used to identify the presence of small-world behaviour in the graph (see below).
This metric has seen many interesting real world applications in the study of the human brain, as well as many other areas, \eg{}\cite{HKB2007}. 

Global graph efficiency \(E_G\) can be defined as:
\begin{equation}\label{eq:efficiency}
    E_G = \frac{1}{N_V(N_V-1)}  \sum_{i \neq j \in V} \frac{1}{dist(i, j)},
\end{equation}
where \(dist\) is the shortest path between node $i$ and $j$.

A key benefit of using \(E_G\) is that it is bounded between zero and 1, making it numerically simpler to compare between graphs.

\subsection*{Node Centrality}\label{sec:centrality}

There are many use cases for which it would be beneficial to measure the relative importance of a given node within the overall graph structure, \eg{} to identify key neurons in a neuronal circuit or assembly.
One such way of measuring this is node centrality, within which there are numerous methods proposed in the literature which measure different aspects of topological structure, \ie{} the underlying graph structure.
Some of these methods originate in the study of web and social networks, with the PageRank algorithm being a famous example as it formed a key part of the early Google search algorithm~\cite{PBMW1999}.
In addition to this, some of the other frequently used centrality measures include Degree, Eigenvector, Closeness and Betweenness~\cite{K2010}. 

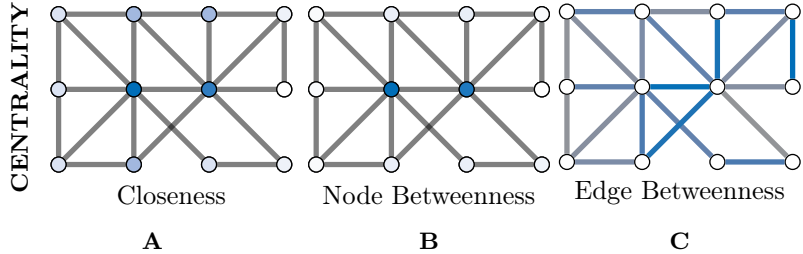
\begin{figure}
    \centering%
    \begin{subfigure}[b][][b]{0.30\linewidth}
        \begin{tikzpicture}
    \node at (-1.75,-0.1) [rotate=90,anchor=north] {\small{\textbf{CENTRALITY}}};
    \begin{scope}[shift={(-1,0)}]
        \GraphInit[vstyle=Simple]
        \SetGraphUnit{1}
        \tikzset{VertexStyle/.append style = {minimum size=5pt, fill=NavyBlue!0}}
        \Vertex{V1} \NO(V1){V2} \SO(V1){V3}
        \EA(V1){V4} \NOEA(V1){V5} \SOEA(V1){V6}
        \EA(V4){V7} \NOEA(V4){V8} \SOEA(V4){V9}
        \EA(V7){V10} \NOEA(V7){V11} \SOEA(V7){V12}
        \AddVertexColor{NavyBlue!13}{V1}
        \AddVertexColor{NavyBlue!13}{V2}q
        \AddVertexColor{NavyBlue!13}{V3}
        \AddVertexColor{NavyBlue!100}{V4}
        \AddVertexColor{NavyBlue!37}{V5}
        \AddVertexColor{NavyBlue!37}{V6}
        \AddVertexColor{NavyBlue!84}{V7}
        \AddVertexColor{NavyBlue!37}{V8}
        \AddVertexColor{NavyBlue!13}{V9}
        \AddVertexColor{NavyBlue!0}{V10}
        \AddVertexColor{NavyBlue!6}{V11}
        \AddVertexColor{NavyBlue!6}{V12}

        \Edge[local=true,lw=2.0pt,style={opacity=0.5}](V1)(V2)
        \Edge[local=true,lw=2.0pt,style={opacity=0.5}](V1)(V3)
        \Edge[local=true,lw=2.0pt,style={opacity=0.5}](V1)(V4)
        \Edge[local=true,lw=2.0pt,style={opacity=0.5}](V2)(V4)
        \Edge[local=true,lw=2.0pt,style={opacity=0.5}](V2)(V5)
        \Edge[local=true,lw=2.0pt,style={opacity=0.5}](V3)(V4)
        \Edge[local=true,lw=2.0pt,style={opacity=0.5}](V3)(V6)
        \Edge[local=true,lw=2.0pt,style={opacity=0.5}](V4)(V5)
        \Edge[local=true,lw=2.0pt,style={opacity=0.5}](V4)(V6)
        \Edge[local=true,lw=2.0pt,style={opacity=0.5}](V4)(V7)
        \Edge[local=true,lw=2.0pt,style={opacity=0.5}](V4)(V9)
        \Edge[local=true,lw=2.0pt,style={opacity=0.5}](V5)(V7)
        \Edge[local=true,lw=2.0pt,style={opacity=0.5}](V5)(V8)
        \Edge[local=true,lw=2.0pt,style={opacity=0.5}](V6)(V7)
        \Edge[local=true,lw=2.0pt,style={opacity=0.5}](V7)(V8)
        \Edge[local=true,lw=2.0pt,style={opacity=0.5}](V7)(V10)
        \Edge[local=true,lw=2.0pt,style={opacity=0.5}](V7)(V11)
        \Edge[local=true,lw=2.0pt,style={opacity=0.5}](V7)(V12)
        \Edge[local=true,lw=2.0pt,style={opacity=0.5}](V8)(V11)
        \Edge[local=true,lw=2.0pt,style={opacity=0.5}](V9)(V12)
        \Edge[local=true,lw=2.0pt,style={opacity=0.5}](V10)(V11)
      \end{scope}%
      \node at (0.5,-1.4) {Closeness};
\end{tikzpicture}%
        \caption{}\label{fig:centrality:closeness}%
    \end{subfigure}%
    \begin{subfigure}[b][][b]{0.25\linewidth}
        \centering%
        \begin{tikzpicture}
    \begin{scope}[shift={(-1,0)}]
        \GraphInit[vstyle=Simple]
        \SetGraphUnit{1}
        \tikzset{VertexStyle/.append style = {minimum size=5pt, fill=NavyBlue!0}}
        \Vertex{V1} \NO(V1){V2} \SO(V1){V3}
        \EA(V1){V4} \NOEA(V1){V5} \SOEA(V1){V6}
        \EA(V4){V7} \NOEA(V4){V8} \SOEA(V4){V9}
        \EA(V7){V10} \NOEA(V7){V11} \SOEA(V7){V12}
        \AddVertexColor{NavyBlue!2}{V1}
        \AddVertexColor{NavyBlue!2}{V2}
        \AddVertexColor{NavyBlue!2}{V3}
        \AddVertexColor{NavyBlue!100}{V4}
        \AddVertexColor{NavyBlue!7}{V5}
        \AddVertexColor{NavyBlue!7}{V6}
        \AddVertexColor{NavyBlue!88}{V7}
        \AddVertexColor{NavyBlue!8}{V8}
        \AddVertexColor{NavyBlue!7}{V9}
        \AddVertexColor{NavyBlue!0}{V10}
        \AddVertexColor{NavyBlue!2}{V11}
        \AddVertexColor{NavyBlue!5}{V12}

        \Edge[local=true,lw=2.0pt,style={opacity=0.5}](V1)(V2)
        \Edge[local=true,lw=2.0pt,style={opacity=0.5}](V1)(V3)
        \Edge[local=true,lw=2.0pt,style={opacity=0.5}](V1)(V4)
        \Edge[local=true,lw=2.0pt,style={opacity=0.5}](V2)(V4)
        \Edge[local=true,lw=2.0pt,style={opacity=0.5}](V2)(V5)
        \Edge[local=true,lw=2.0pt,style={opacity=0.5}](V3)(V4)
        \Edge[local=true,lw=2.0pt,style={opacity=0.5}](V3)(V6)
        \Edge[local=true,lw=2.0pt,style={opacity=0.5}](V4)(V5)
        \Edge[local=true,lw=2.0pt,style={opacity=0.5}](V4)(V6)
        \Edge[local=true,lw=2.0pt,style={opacity=0.5}](V4)(V7)
        \Edge[local=true,lw=2.0pt,style={opacity=0.5}](V4)(V9)
        \Edge[local=true,lw=2.0pt,style={opacity=0.5}](V5)(V7)
        \Edge[local=true,lw=2.0pt,style={opacity=0.5}](V5)(V8)
        \Edge[local=true,lw=2.0pt,style={opacity=0.5}](V6)(V7)
        \Edge[local=true,lw=2.0pt,style={opacity=0.5}](V7)(V8)
        \Edge[local=true,lw=2.0pt,style={opacity=0.5}](V7)(V10)
        \Edge[local=true,lw=2.0pt,style={opacity=0.5}](V7)(V11)
        \Edge[local=true,lw=2.0pt,style={opacity=0.5}](V7)(V12)
        \Edge[local=true,lw=2.0pt,style={opacity=0.5}](V8)(V11)
        \Edge[local=true,lw=2.0pt,style={opacity=0.5}](V9)(V12)
        \Edge[local=true,lw=2.0pt,style={opacity=0.5}](V10)(V11)
      \end{scope}%
      \node at (0.5,-1.4) {Node Betweenness};
\end{tikzpicture}%
        \caption{}\label{fig:centrality:vbetweenness}%
    \end{subfigure}%
    \begin{subfigure}[b][][b]{0.25\linewidth}
        \centering%
        \begin{tikzpicture}
    \begin{scope}[shift={(-1,0)}]
        \GraphInit[vstyle=Simple]
        \SetGraphUnit{1}
        \tikzset{VertexStyle/.append style = {minimum size=5pt, fill=NavyBlue!0}}
        \Vertex{V1} \NO(V1){V2} \SO(V1){V3}
        \EA(V1){V4} \NOEA(V1){V5} \SOEA(V1){V6}
        \EA(V4){V7} \NOEA(V4){V8} \SOEA(V4){V9}
        \EA(V7){V10} \NOEA(V7){V11} \SOEA(V7){V12}

        \Edge[local=true,
              lw=2.0pt,
              color=NavyBlue!5!Gray,
              style={opacity=1}](V1)(V2)
        \Edge[local=true,
              lw=2.0pt,
              color=NavyBlue!5!Gray,
              style={opacity=1}](V1)(V3)
        \Edge[local=true,
              lw=2.0pt,
              color=NavyBlue!58!Gray,
              style={opacity=1}](V1)(V4)
        \Edge[local=true,
              lw=2.0pt,
              color=NavyBlue!18!Gray,
              style={opacity=1}](V2)(V4)
        \Edge[local=true,
              lw=2.0pt,
              color=NavyBlue!58!Gray,
              style={opacity=1}](V2)(V5)
        \Edge[local=true,
              lw=2.0pt,
              color=NavyBlue!18!Gray,
              style={opacity=1}](V3)(V4)
        \Edge[local=true,
              lw=2.0pt,
              color=NavyBlue!33!Gray,
              style={opacity=1}](V3)(V6)
        \Edge[local=true,
              lw=2.0pt,
              color=NavyBlue!33!Gray,
              style={opacity=1}](V4)(V5)
        \Edge[local=true,
              lw=2.0pt,
              color=NavyBlue!71!Gray,
              style={opacity=1}](V4)(V6)
        \Edge[local=true,
              lw=2.0pt,
              color=NavyBlue!96!Gray,
              style={opacity=1}](V4)(V7)
        \Edge[local=true,
              lw=2.0pt,
              color=NavyBlue!56!Gray,
              style={opacity=1}](V4)(V9)
        \Edge[local=true,
              lw=2.0pt,
              color=NavyBlue!56!Gray,
              style={opacity=1}](V5)(V7)
        \Edge[local=true,
              lw=2.0pt,
              color=NavyBlue!22!Gray,
              style={opacity=1}](V5)(V8)
        \Edge[local=true,
              lw=2.0pt,
              color=NavyBlue!87!Gray,
              style={opacity=1}](V6)(V7)
        \Edge[local=true,
              lw=2.0pt,
              color=NavyBlue!89!Gray,
              style={opacity=1}](V7)(V8)
        \Edge[local=true,
              lw=2.0pt,
              color=NavyBlue!22!Gray,
              style={opacity=1}](V7)(V10)
        \Edge[local=true,
              lw=2.0pt,
              color=NavyBlue!27!Gray,
              style={opacity=1}](V7)(V11)
        \Edge[local=true,
              lw=2.0pt,
              color=NavyBlue!0!Gray,
              style={opacity=1}](V7)(V12)
        \Edge[local=true,
              lw=2.0pt,
              color=NavyBlue!71!Gray,
              style={opacity=1}](V9)(V12)
        \Edge[local=true,
              lw=2.0pt,
              color=NavyBlue!100!Gray,
              style={opacity=1}](V10)(V11)
        \Edge[local=true,
              lw=2.0pt,
              color=NavyBlue!60!Gray,
              style={opacity=1}](V8)(V11)
      \end{scope}%
      \node at (0.5,-1.4) {Edge Betweenness};
\end{tikzpicture}%
        \caption{}\label{fig:centrality:ebetweenness}%
    \end{subfigure}%
    \caption{Centrality measures indicate the relative importance of a node within a graph.
        (\textbf{A}) Closeness centrality gives a relative importance based on the average shortest patch between a node and all other nodes. The smaller the average the more important the node (darker blue) and vice versa (whiter).
        (\textbf{B}) Betweenness centrality is similar but clearly separate the central two nodes (dark blue) as much more important than the other nodes (which appear white). 
        (\textbf{C}) Centrality can also be applied to edges instead of nodes; here more blue indicates a more central role in information flow for an edge.
    }\label{fig:centrality}%
\end{figure}

We will explore Closeness and Betweenness centrality in greater detail.
Closeness centrality computes a nodes importance by measuring its average `distance' to all other nodes in the graph, where the shortest path between two nodes is used as the distance metric.
So for a given node \(i \in V\) from \(G(V, E)\), its Closeness centrality would be given as
\begin{equation}\label{eq:cc}
    CC_i = \frac{1}{\sum\limits_{j\in V} dist(i, j)}~,
\end{equation}
where \(dist\) is the shortest path between \(i\) and \(j\).
This is visualised in \cref{fig:centrality:closeness}, where the two nodes in the dark blue colour have the highest Closeness centrality score as they posses the lowest overall average shortest path length to the other nodes.

Additionally, ~\cref{fig:centrality:vbetweenness} and ~\cref{fig:centrality:ebetweenness} demonstrates both node and edge Betweenness centrality measures respectively. Betweenness centrality exploits the concept of shortest paths (discussed earlier) to argue that nodes through which a greater volume of shortest paths pass through, are of greater importance in the graph \cite{F1977}. Therefore, nodes with a high value of Betweenness centrality can be seen as controlling the information flow between other nodes in the graph. Edge Betweenness is a measure, which analogous to its node counterpart, counts the number of shortest paths which travel along each edge~\cite{N2010}.

\subsection*{Graph Motifs}\label{sec:motifs}

A more complex measure of graph local topology is that of a motif, a small and reoccurring local pattern of connectivity between nodes in a graph. 
It is argued that one can consider these motifs to be the building block of more complex patterns of connectivity within graphs~\cite{SMMA2002}.
Indeed, the type and frequency of certain motifs can even be used for tasks such as graph comparison~\cite{MP2008} and graph classification~\cite{HWAPST2003}.

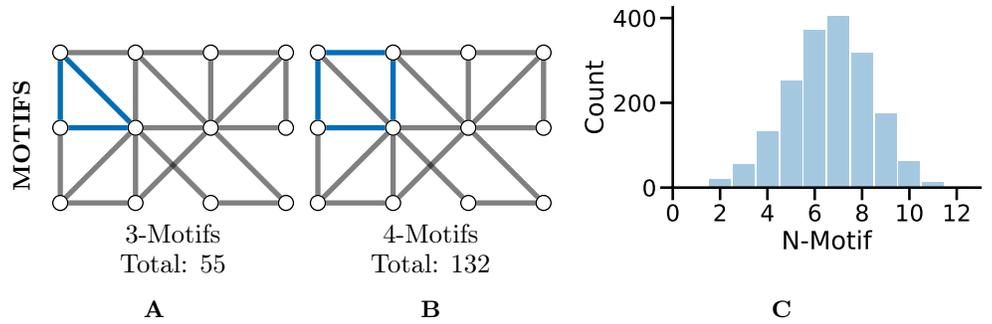
\begin{figure}
    \centering%
    \begin{subfigure}[b][][b]{0.3\linewidth}
        \begin{tikzpicture}
    \node at (-1.75,-0.1) [rotate=90,anchor=north] {\small{\textbf{MOTIFS}}};
    \begin{scope}[shift={(-1,0)}]
        \GraphInit[vstyle=Simple]
        \SetGraphUnit{1}
        \tikzset{VertexStyle/.append style = {minimum size=5pt, fill=NavyBlue!0}}
        \Vertex{V1} \NO(V1){V2} \SO(V1){V3}
        \EA(V1){V4} \NOEA(V1){V5} \SOEA(V1){V6}
        \EA(V4){V7} \NOEA(V4){V8} \SOEA(V4){V9}
        \EA(V7){V10} \NOEA(V7){V11} \SOEA(V7){V12}

        \Edge[local=true,lw=2.0pt,color=NavyBlue](V1)(V2)
        \Edge[local=true,lw=2.0pt,style={opacity=0.5}](V1)(V3)
        \Edge[local=true,lw=2.0pt,color=NavyBlue](V1)(V4)
        \Edge[local=true,lw=2.0pt,color=NavyBlue](V2)(V4)
        \Edge[local=true,lw=2.0pt,style={opacity=0.5}](V2)(V5)
        \Edge[local=true,lw=2.0pt,style={opacity=0.5}](V3)(V4)
        \Edge[local=true,lw=2.0pt,style={opacity=0.5}](V3)(V6)
        \Edge[local=true,lw=2.0pt,style={opacity=0.5}](V4)(V5)
        \Edge[local=true,lw=2.0pt,style={opacity=0.5}](V4)(V6)
        \Edge[local=true,lw=2.0pt,style={opacity=0.5}](V4)(V7)
        \Edge[local=true,lw=2.0pt,style={opacity=0.5}](V4)(V9)
        \Edge[local=true,lw=2.0pt,style={opacity=0.5}](V5)(V7)
        \Edge[local=true,lw=2.0pt,style={opacity=0.5}](V5)(V8)
        \Edge[local=true,lw=2.0pt,style={opacity=0.5}](V6)(V7)
        \Edge[local=true,lw=2.0pt,style={opacity=0.5}](V7)(V8)
        \Edge[local=true,lw=2.0pt,style={opacity=0.5}](V7)(V10)
        \Edge[local=true,lw=2.0pt,style={opacity=0.5}](V7)(V11)
        \Edge[local=true,lw=2.0pt,style={opacity=0.5}](V7)(V12)
        \Edge[local=true,lw=2.0pt,style={opacity=0.5}](V8)(V11)
        \Edge[local=true,lw=2.0pt,style={opacity=0.5}](V9)(V12)
        \Edge[local=true,lw=2.0pt,style={opacity=0.5}](V10)(V11)
      \end{scope}%
      \node at (0.5,-1.4) {3-Motifs};
      \node at (0.5,-1.8) {Total: 55};
\end{tikzpicture}%
        \caption{}\label{fig:motifs:3}%
    \end{subfigure}%
    \begin{subfigure}[b][][b]{0.25\linewidth}
        \centering%
        \begin{tikzpicture}
    \begin{scope}[shift={(-1,0)}]
        \GraphInit[vstyle=Simple]
        \SetGraphUnit{1}
        \tikzset{VertexStyle/.append style = {minimum size=5pt, fill=NavyBlue!0}}
        \Vertex{V1} \NO(V1){V2} \SO(V1){V3}
        \EA(V1){V4} \NOEA(V1){V5} \SOEA(V1){V6}
        \EA(V4){V7} \NOEA(V4){V8} \SOEA(V4){V9}
        \EA(V7){V10} \NOEA(V7){V11} \SOEA(V7){V12}

        \Edge[local=true,lw=2.0pt,color=NavyBlue](V1)(V2)
        \Edge[local=true,lw=2.0pt,style={opacity=0.5}](V1)(V3)
        \Edge[local=true,lw=2.0pt,color=NavyBlue](V1)(V4)
        \Edge[local=true,lw=2.0pt,style={opacity=0.5}](V2)(V4)
        \Edge[local=true,lw=2.0pt,color=NavyBlue](V2)(V5)
        \Edge[local=true,lw=2.0pt,style={opacity=0.5}](V3)(V4)
        \Edge[local=true,lw=2.0pt,style={opacity=0.5}](V3)(V6)
        \Edge[local=true,lw=2.0pt,color=NavyBlue](V4)(V5)
        \Edge[local=true,lw=2.0pt,style={opacity=0.5}](V4)(V6)
        \Edge[local=true,lw=2.0pt,style={opacity=0.5}](V4)(V7)
        \Edge[local=true,lw=2.0pt,style={opacity=0.5}](V4)(V9)
        \Edge[local=true,lw=2.0pt,style={opacity=0.5}](V5)(V7)
        \Edge[local=true,lw=2.0pt,style={opacity=0.5}](V5)(V8)
        \Edge[local=true,lw=2.0pt,style={opacity=0.5}](V6)(V7)
        \Edge[local=true,lw=2.0pt,style={opacity=0.5}](V7)(V8)
        \Edge[local=true,lw=2.0pt,style={opacity=0.5}](V7)(V10)
        \Edge[local=true,lw=2.0pt,style={opacity=0.5}](V7)(V11)
        \Edge[local=true,lw=2.0pt,style={opacity=0.5}](V7)(V12)
        \Edge[local=true,lw=2.0pt,style={opacity=0.5}](V8)(V11)
        \Edge[local=true,lw=2.0pt,style={opacity=0.5}](V9)(V12)
        \Edge[local=true,lw=2.0pt,style={opacity=0.5}](V10)(V11)
      \end{scope}%
      \node at (0.5,-1.4) {4-Motifs};
      \node at (0.5,-1.8) {Total: 132};
\end{tikzpicture}%
        \caption{}\label{fig:motifs:4}%
    \end{subfigure}%
    \begin{subfigure}[b][][b]{0.45\linewidth}
        \centering%
        \includegraphics[width=\linewidth]{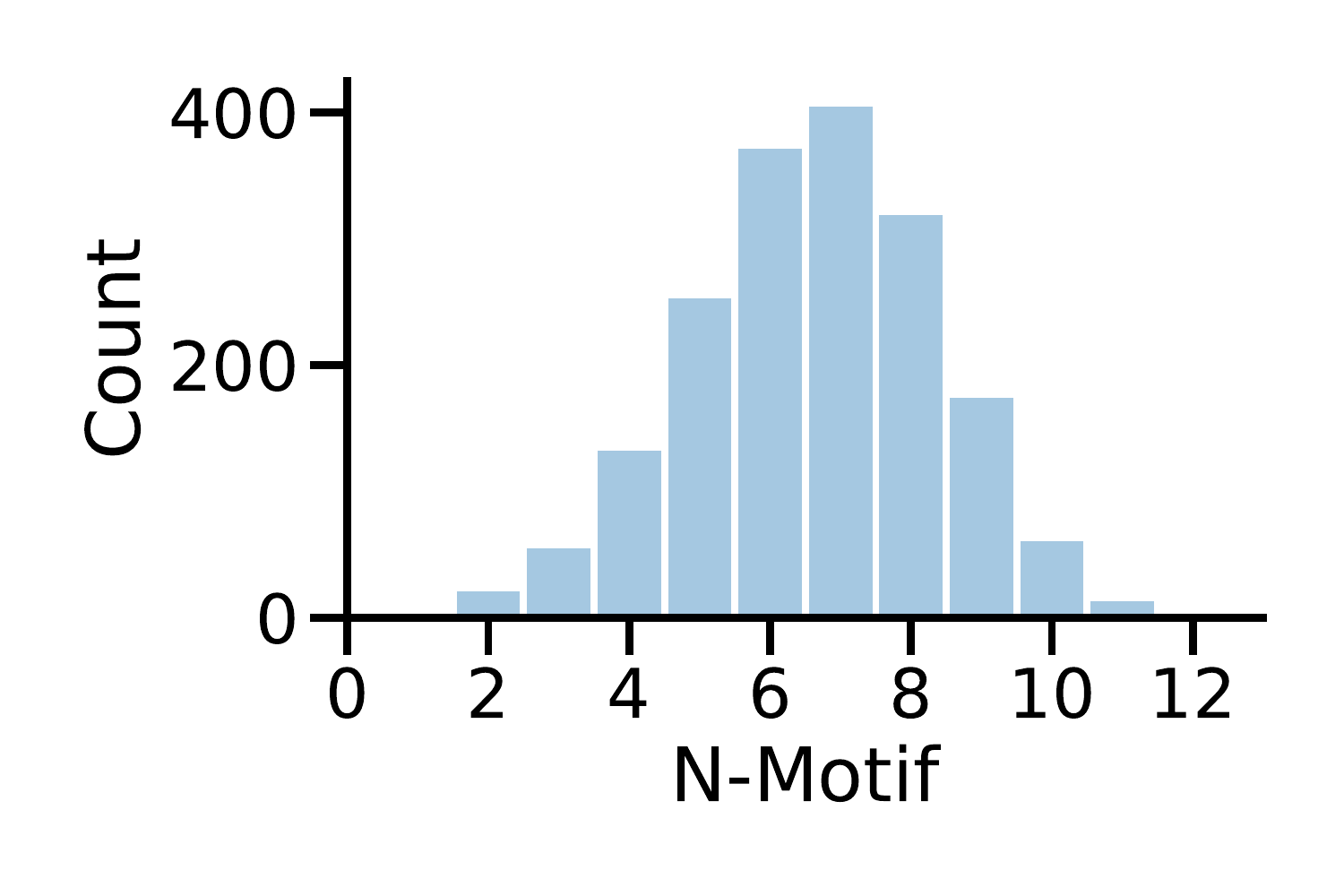}%
        \caption{}\label{fig:motifs:hist}%
    \end{subfigure}%
    \caption{Motifs represent repeating local topological patterns in the graph.
        (\textbf{A}) An example 3-motif in blue; there are a total of 55 3-motifs in this graph.
        (\textbf{B}) An equivalent 4-motif in blue; there are 132 4-motifs in this graph. 
        (\textbf{C}) Histograms of graph motif counts can be used to create a signature for graphs that can then be compared.
    }\label{fig:motifs}%
\end{figure}

Perhaps the most fundamental motif is that of the triangle (see~\cref{fig:motifs:3}), a series of three nodes where an edge is present between all the nodes.
A similar motif comprised of four nodes is highlighted in~\cref{fig:motifs:4}.
The study of motifs in graphs has proved popular in fMRI studies where distributions of motifs have been used to separate clinical cases, \eg{}~\cite{JZCQM2014, MATB2018}.

In~\cite{WF2019}, the authors used motif analysis in a directed representation of mitral cell and interneuron functional connectivity in the olfactory bulb of zebrafish larvae.
They found that motifs with reciprocal edges were over-represented and mediate inhibition between neurons with similar tuning.
The resultant suppression of redundancy, inferred from theoretical models and tested through selective manipulations of simulations, was necessary and sufficient to reproduce a fundamental computation known as `whitening'.

\subsection*{Graph Clustering Coefficient}\label{sec:cluster}

\begin{figure}
    \centering%
    \begin{subfigure}[b][][b]{0.3\linewidth}
        \begin{tikzpicture}
    \node at (-1.75,-0.1) [rotate=90,anchor=north] {\small{\textbf{CLUSTERING}}};
    \begin{scope}[shift={(-1,0)}]
        \GraphInit[vstyle=Simple]
        \SetGraphUnit{1}
        \tikzset{VertexStyle/.append style = {minimum size=5pt, fill=NavyBlue!0}}
        \Vertex{V1} \NO(V1){V2} \SO(V1){V3}
        \EA(V1){V4} \NOEA(V1){V5} \SOEA(V1){V6}
        \EA(V4){V7} \NOEA(V4){V8} \SOEA(V4){V9}
        \EA(V7){V10} \NOEA(V7){V11} \SOEA(V7){V12}

        \Edge[local=true,lw=2.0pt,color=NavyBlue](V1)(V2)
        \Edge[local=true,lw=2.0pt,style={opacity=0.5}](V1)(V3)
        \Edge[local=true,lw=2.0pt,color=NavyBlue](V1)(V4)
        \Edge[local=true,lw=2.0pt,color=NavyBlue](V2)(V4)
        \Edge[local=true,lw=2.0pt,style={opacity=0.5}](V2)(V5)
        \Edge[local=true,lw=2.0pt,style={opacity=0.5}](V3)(V4)
        \Edge[local=true,lw=2.0pt,style={opacity=0.5}](V3)(V6)
        \Edge[local=true,lw=2.0pt,style={opacity=0.5}](V4)(V5)
        \Edge[local=true,lw=2.0pt,style={opacity=0.5}](V4)(V6)
        \Edge[local=true,lw=2.0pt,style={opacity=0.5}](V4)(V7)
        \Edge[local=true,lw=2.0pt,style={opacity=0.5}](V4)(V9)
        \Edge[local=true,lw=2.0pt,style={opacity=0.5}](V5)(V7)
        \Edge[local=true,lw=2.0pt,style={opacity=0.5}](V5)(V8)
        \Edge[local=true,lw=2.0pt,style={opacity=0.5}](V6)(V7)
        \Edge[local=true,lw=2.0pt,style={opacity=0.5}](V7)(V8)
        \Edge[local=true,lw=2.0pt,color=ForestGreen](V7)(V10)
        \Edge[local=true,lw=2.0pt,style={opacity=0.5}](V7)(V11)
        \Edge[local=true,lw=2.0pt,color=ForestGreen](V7)(V12)
        \Edge[local=true,lw=2.0pt,style={opacity=0.5}](V8)(V11)
        \Edge[local=true,lw=2.0pt,style={opacity=0.5}](V9)(V12)
        \Edge[local=true,lw=2.0pt,style={opacity=0.5}](V10)(V11)
        \Edge[local=true,lw=2.0pt,color=ForestGreen,style={dashed}](V10)(V12)
      \end{scope}%
      \node at (0.5,-1.4) [NavyBlue] {Closed Triplet};
      \node at (0.5,-1.8) [ForestGreen] {Open Triplet};
\end{tikzpicture}%
        \caption{}\label{fig:clustering:triplets}%
    \end{subfigure}%
    \begin{subfigure}[b][][b]{0.25\linewidth}
        \centering%
        \begin{tikzpicture}
    \begin{scope}[shift={(-1,0)}]
        \GraphInit[vstyle=Simple]
        \SetGraphUnit{1}
        \tikzset{VertexStyle/.append style = {minimum size=5pt, fill=NavyBlue!0}}
        \Vertex{V1} \NO(V1){V2} \SO(V1){V3}
        \EA(V1){V4} \NOEA(V1){V5} \SOEA(V1){V6}
        \EA(V4){V7} \NOEA(V4){V8} \SOEA(V4){V9}
        \EA(V7){V10} \NOEA(V7){V11} \SOEA(V7){V12}
        \AddVertexColor{NavyBlue!67}{V1}
        \AddVertexColor{NavyBlue!67}{V2}
        \AddVertexColor{NavyBlue!67}{V3}
        \AddVertexColor{NavyBlue!25}{V4}
        \AddVertexColor{NavyBlue!67}{V5}
        \AddVertexColor{NavyBlue!67}{V6}
        \AddVertexColor{NavyBlue!24}{V7}
        \AddVertexColor{NavyBlue!67}{V8}
        \AddVertexColor{NavyBlue!0}{V9}
        \AddVertexColor{NavyBlue!1}{V10}
        \AddVertexColor{NavyBlue!67}{V11}
        \AddVertexColor{NavyBlue!0}{V12}

        \Edge[local=true,lw=2.0pt,style={opacity=0.5}](V1)(V2)
        \Edge[local=true,lw=2.0pt,style={opacity=0.5}](V1)(V3)
        \Edge[local=true,lw=2.0pt,style={opacity=0.5}](V1)(V4)
        \Edge[local=true,lw=2.0pt,style={opacity=0.5}](V2)(V4)
        \Edge[local=true,lw=2.0pt,style={opacity=0.5}](V2)(V5)
        \Edge[local=true,lw=2.0pt,style={opacity=0.5}](V3)(V4)
        \Edge[local=true,lw=2.0pt,style={opacity=0.5}](V3)(V6)
        \Edge[local=true,lw=2.0pt,style={opacity=0.5}](V4)(V5)
        \Edge[local=true,lw=2.0pt,style={opacity=0.5}](V4)(V6)
        \Edge[local=true,lw=2.0pt,style={opacity=0.5}](V4)(V7)
        \Edge[local=true,lw=2.0pt,style={opacity=0.5}](V4)(V9)
        \Edge[local=true,lw=2.0pt,style={opacity=0.5}](V5)(V7)
        \Edge[local=true,lw=2.0pt,style={opacity=0.5}](V5)(V8)
        \Edge[local=true,lw=2.0pt,style={opacity=0.5}](V6)(V7)
        \Edge[local=true,lw=2.0pt,style={opacity=0.5}](V7)(V8)
        \Edge[local=true,lw=2.0pt,style={opacity=0.5}](V7)(V10)
        \Edge[local=true,lw=2.0pt,style={opacity=0.5}](V7)(V11)
        \Edge[local=true,lw=2.0pt,style={opacity=0.5}](V7)(V12)
        \Edge[local=true,lw=2.0pt,style={opacity=0.5}](V8)(V11)
        \Edge[local=true,lw=2.0pt,style={opacity=0.5}](V9)(V12)
        \Edge[local=true,lw=2.0pt,style={opacity=0.5}](V10)(V11)
      \end{scope}%
      \node at (0.5,-1.4) [NavyBlue] {Local \(\in [0,1]\)};
      \node at (0.5,-1.8) [ForestGreen] {Global: \(0.37\)};
\end{tikzpicture}%
        \caption{}\label{fig:clustering:local}%
    \end{subfigure}%
    \caption{Clustering provides another measure of connectivity and structure in a graph.
        (\textbf{A}) Based the number of closed (blue) and open (green) triplets, the clustering coefficient can be calculated locally for every node.
        (\textbf{B}) Local clustering coefficients for nodes range from zero (white) to one (blue) and may vary a lot from the global (mean) clustering coefficient.
    }\label{fig:clustering}%
\end{figure}
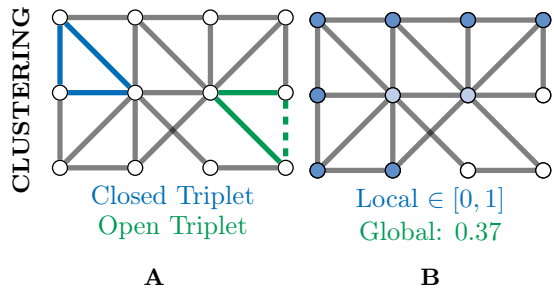

A further measure of local connectivity within a graph is that of the clustering coefficient.
At the level of individual nodes, the clustering coefficient gives a measure of how connected that node's neighbourhood is within itself.
For example, in~\cref{fig:clustering:local}, the nodes coloured in white have a low local clustering coefficient as their neighbourhoods are not densely connected. 
More concretely, for a given node \(v\), the clustering coefficient determines the fraction of one-hop neighbours of \(v\) which are themselves connected via an edge,
\begin{equation}
C_v = \frac{\text{number of closed triplets}}{\text{number of all triplets}}~,
\end{equation}
where triplets refers to all possible combinations of three of neighbours of \(v\), both open and closed~\cite{N2010}.
An example of a closed and open triplet in a graph is illustrated in~\cref{fig:clustering:triplets}.

To produce a single metric representing the overall level of connectivity within a graph, the global clustering coefficient is used $C_G$. This is simply the mean local clustering coefficient over all nodes and can be computed as:
\begin{equation}
    C_G = \frac{1}{N} \sum_{v\in V} C_v~.
\end{equation}

Clustering of a neuronal graph can be used to show differences in the functional organisation of network with graphs of high average clustering coefficient thought to be better at local information integration and robust to disruption.
For example,~\cite{BHGB2019} showed that the clustering coefficient of whole-brain graphs in wild type fish and a depression-like mutant (\emph{gr\textsuperscript{s357}}) differ and, importantly this can be restored with the application of anti-depressant drugs.
This change in local connectivity imply that depression increases local brain segregation reducing local information transfer efficiency.

\subsection*{Graph Communities}\label{sec:communities}

Community detection in graphs is a large area of interest within the literature and could be an entire review within itself.
As such, we will outline the major concepts here and direct interested readers towards more in-depth reviews on community detection such as \cite{F2010} and \cite{YAT2016}.

Fundamentally, one can view communities as partitions or clusters of nodes within a graph, where members of the clusters contain similar (as defined by some metric) nodes.
These global communities differ from the local, more pattern-focussed, graph motifs previously discussed.
Community structures relate to specialisations within networks, \eg{} a social media graph might see community structures relating to shared hobbies or interests.
In graphs of brains, high-levels of community structure could indicate functional specialisation~\cite{BS2009}.

\begin{figure}
    \centering%
    \begin{subfigure}[b][][b]{0.3\linewidth}
        \begin{tikzpicture}
    \node at (-1.75,-0.1) [rotate=90,anchor=north] {\small{\textbf{COMMUNITIES}}};
    \begin{scope}[shift={(-1,0)}]
        \GraphInit[vstyle=Simple]
        \SetGraphUnit{1}
        \tikzset{VertexStyle/.append style = {minimum size=5pt, fill=NavyBlue!0}}
        \Vertex{V1} \NO(V1){V2} \SO(V1){V3}
        \EA(V1){V4} \NOEA(V1){V5} \SOEA(V1){V6}
        \EA(V4){V7} \NOEA(V4){V8} \SOEA(V4){V9}
        \EA(V7){V10} \NOEA(V7){V11} \SOEA(V7){V12}
        \AddVertexColor{NavyBlue}{V1}
        \AddVertexColor{NavyBlue}{V2}
        \AddVertexColor{NavyBlue}{V3}
        \AddVertexColor{NavyBlue}{V4}
        \AddVertexColor{NavyBlue}{V5}
        \AddVertexColor{NavyBlue}{V6}
        \AddVertexColor{ForestGreen}{V7}
        \AddVertexColor{ForestGreen}{V8}
        \AddVertexColor{ForestGreen}{V9}
        \AddVertexColor{ForestGreen}{V10}
        \AddVertexColor{ForestGreen}{V11}
        \AddVertexColor{ForestGreen}{V12}

        \Edge[local=true,lw=2.0pt,style={opacity=0.5}](V1)(V2)
        \Edge[local=true,lw=2.0pt,style={opacity=0.5}](V1)(V3)
        \Edge[local=true,lw=2.0pt,style={opacity=0.5}](V1)(V4)
        \Edge[local=true,lw=2.0pt,style={opacity=0.5}](V2)(V4)
        \Edge[local=true,lw=2.0pt,style={opacity=0.5}](V2)(V5)
        \Edge[local=true,lw=2.0pt,style={opacity=0.5}](V3)(V4)
        \Edge[local=true,lw=2.0pt,style={opacity=0.5}](V3)(V6)
        \Edge[local=true,lw=2.0pt,style={opacity=0.5}](V4)(V5)
        \Edge[local=true,lw=2.0pt,style={opacity=0.5}](V4)(V6)
        \Edge[local=true,lw=2.0pt,style={opacity=0.5}](V4)(V7)
        \Edge[local=true,lw=2.0pt,style={opacity=0.5}](V4)(V9)
        \Edge[local=true,lw=2.0pt,style={opacity=0.5}](V5)(V7)
        \Edge[local=true,lw=2.0pt,style={opacity=0.5}](V7)(V8)
        \Edge[local=true,lw=2.0pt,style={opacity=0.5}](V7)(V9)
        \Edge[local=true,lw=2.0pt,style={opacity=0.5}](V7)(V10)
        \Edge[local=true,lw=2.0pt,style={opacity=0.5}](V7)(V11)
        \Edge[local=true,lw=2.0pt,style={opacity=0.5}](V7)(V12)
        \Edge[local=true,lw=2.0pt,style={opacity=0.5}](V8)(V11)
        \Edge[local=true,lw=2.0pt,style={opacity=0.5}](V9)(V12)
        \Edge[local=true,lw=2.0pt,style={opacity=0.5}](V10)(V11)
        \Edge[local=true,lw=2.0pt,style={opacity=0.5}](V10)(V12)
      \end{scope}%
      \node at (0.5,-1.4) [NavyBlue] {Non-overlapping A};
      \node at (0.5,-2.2) [ForestGreen] {Non-overlapping B};
\end{tikzpicture}%
        \caption{}\label{fig:communities:nonoverlapping}%
    \end{subfigure}%
    \begin{subfigure}[b][][b]{0.25\linewidth}
        \centering%
        \begin{tikzpicture}
    \begin{scope}[shift={(-1,0)}]
        \GraphInit[vstyle=Simple]
        \SetGraphUnit{1}
        \tikzset{VertexStyle/.append style = {minimum size=5pt, fill=NavyBlue!0}}
        \Vertex{V1} \NO(V1){V2} \SO(V1){V3}
        \EA(V1){V4} \NOEA(V1){V5} \SOEA(V1){V6}
        \EA(V4){V7} \NOEA(V4){V8} \SOEA(V4){V9}
        \EA(V7){V10} \NOEA(V7){V11} \SOEA(V7){V12}
        \AddVertexColor{NavyBlue}{V1}
        \AddVertexColor{NavyBlue}{V2}
        \AddVertexColor{NavyBlue}{V3}
        \AddVertexColor{Black}{V4}
        \AddVertexColor{NavyBlue}{V5}
        \AddVertexColor{NavyBlue}{V6}
        \AddVertexColor{Black}{V7}
        \AddVertexColor{ForestGreen}{V8}
        \AddVertexColor{ForestGreen}{V9}
        \AddVertexColor{ForestGreen}{V10}
        \AddVertexColor{ForestGreen}{V11}
        \AddVertexColor{ForestGreen}{V12}

        \Edge[local=true,lw=2.0pt,style={opacity=0.5}](V1)(V2)
        \Edge[local=true,lw=2.0pt,style={opacity=0.5}](V1)(V3)
        \Edge[local=true,lw=2.0pt,style={opacity=0.5}](V1)(V4)
        \Edge[local=true,lw=2.0pt,style={opacity=0.5}](V2)(V4)
        \Edge[local=true,lw=2.0pt,style={opacity=0.5}](V2)(V5)
        \Edge[local=true,lw=2.0pt,style={opacity=0.5}](V3)(V4)
        \Edge[local=true,lw=2.0pt,style={opacity=0.5}](V3)(V6)
        \Edge[local=true,lw=2.0pt,style={opacity=0.5}](V4)(V5)
        \Edge[local=true,lw=2.0pt,style={opacity=0.5}](V4)(V6)
        \Edge[local=true,lw=2.0pt,style={opacity=0.5}](V4)(V7)
        \Edge[local=true,lw=2.0pt,style={opacity=0.5}](V4)(V9)
        \Edge[local=true,lw=2.0pt,style={opacity=0.5}](V5)(V7)
        \Edge[local=true,lw=2.0pt,style={opacity=0.5}](V7)(V8)
        \Edge[local=true,lw=2.0pt,style={opacity=0.5}](V7)(V9)
        \Edge[local=true,lw=2.0pt,style={opacity=0.5}](V7)(V10)
        \Edge[local=true,lw=2.0pt,style={opacity=0.5}](V7)(V11)
        \Edge[local=true,lw=2.0pt,style={opacity=0.5}](V7)(V12)
        \Edge[local=true,lw=2.0pt,style={opacity=0.5}](V8)(V11)
        \Edge[local=true,lw=2.0pt,style={opacity=0.5}](V9)(V12)
        \Edge[local=true,lw=2.0pt,style={opacity=0.5}](V10)(V11)
        \Edge[local=true,lw=2.0pt,style={opacity=0.5}](V10)(V12)
      \end{scope}%
      \node at (0.5,-1.4) [NavyBlue] {Non-overlapping A};
      \node at (0.5,-1.8) [Black] {Overlapping A \& B};
      \node at (0.5,-2.2) [ForestGreen] {Non-overlapping B};
\end{tikzpicture}%
        \caption{}\label{fig:communities:overlapping}%
    \end{subfigure}%
    \caption{Community detection provides global clustering that can be either non-overlapping or overlapping.
        (\textbf{A}) Non-overlapping communities assign each node to a community (blue or green) based on the choice of metric, often relating to number of connections.
        (\textbf{B}) Over-lapping communities can assign a node to more than one community (black nodes) if they contribute to multiple communities.
    }\label{fig:communities}%
\end{figure}
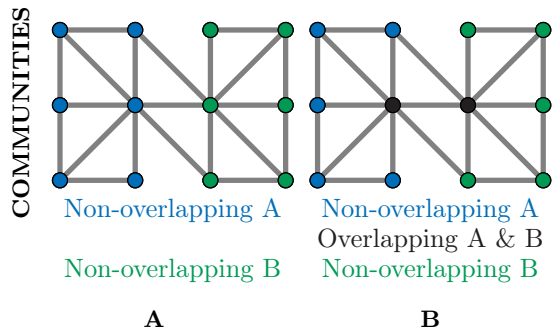

Community detection algorithms can broadly be split into those which produce overlapping communities and those that result in non-overlapping communities~\cite{DZSL2016}.
In a non-overlapping community each node belongs to only one community and, as such, could be used to separate a neuronal graph into distinct regions, \eg{} regions of the brain or layers of the tectum.
This can be seen in ~\cref{fig:communities:nonoverlapping}, where nodes in the graph belong to exactly one community. 
In an overlapping community, nodes may belong to multiple communities and, as such, could be used to identify neural circuits or assemblies where neurons may contribute to multiple pathways~\cite{XKS2013}. This can be seen in ~\cref{fig:communities:overlapping}, where the nodes coloured in black belong to both communities. 

One of the most frequently used approaches for non-overlapping community detection in graphs is that of spectral clustering~\cite{WD2010}.
Here the eigenvectors and eigenvalues of the graph Laplacian matrix are used to detect connectivity based communities.
The distribution of eigenvalues is indicative the total number of clusters within the graph, and the eigenvectors show how to partition the nodes into their respective clusters.

Many approaches for determining communities exploit the concept of modularity to produce their results~\cite{DP2016}.
As this concept is frequently explored in conjunction with biological networks, it is explored in greater depth in the following section.

\subsubsection*{Graph Modularity}\label{sec:modularity}

Strongly linked to the concept of measuring community structure within a graph, is the idea of modularity.
The modularity of a graph is a more fundamental measure of the strength of interconnectivity between communities (or modules as they are commonly known in the modularity literature) of nodes~\cite{N2006}.
Whilst there are different measures of modularity, the majority of them aim to partition a graph in such a way that the intra-community edges are maximized, whilst the number of inter-community edges are minimised~\cite{BNB2017}.
Interestingly, it has been observed that many biological networks, including networks taken from animal brains, display a high degree of modularity \cite{HHLM1999}, perhaps indicative of functional circuits of neurons within the brain.

Modules in a graph confer robustness to networks whilst allowing for specialised processing.
In the mouse auditory cortex it has been shown that neuronal graphs exhibit hierarchically modular structures~\cite{BWAetal2019}.

In zebrafish, brains of `depressed' fish (\emph{gr\textsuperscript{s357}} mutants) show an increased modularity compared to wild-type, which could be restored with anti-depressant drugs~\cite{BHGB2019}.
The combination of reduced clustering coefficient (see above) but increased modularity implies that, functionally, the brain is much less structured and organised in the disease case with more isolated communities of networks and reduced long-range communication.
In \cite{Kha2019}, the author used modularity, amongst other metrics, to compare looming-selective networks in the Xenopus tectum through development and with a range of computational models.

By using spectral clustering and maximising a modularity metric it is also possible extract ensembles of strongly connections neurons, \ie{} neuronal subgraphs.
In~\cite{Kha2019}, the author did this for neurons in the optic tectum of xenopus tadpoles responding to looming stimuli.
They showed that although the number of neuronal subgraphs did not significantly vary at different developmental stages, these neuronal subgraphs were spatially localised and became more distinct throughout development.
This shows reorganisation and refinement of looming-selective neuronal subgraphs within the optic tectum, possibly representing the weakening of functional connections not required for this type of neural computation.

\subsection*{One Metric to Rule Them All?}

Many of these commonly used metrics relate to `clustering', `connectivity' and `organisation' of the graph structure.
One question the reader might ask, is ``If there are so many measures of connectivity, which one do I pick?'' or, later in the analysis process, ``Why do different clustering/community algorithms give me markedly different results?''.
In truth, a change in one particular metric could be due to a variety of changes in the underlying graph~\cite{CSSB2019}.
As such, the answer to both of those questions depends on the hypothesis, experiment and assumptions for that particular scenario.

Scientists who are interested in exploring neuronal graphs for calcium imaging are in luck --- not only is there a large body of technical mathematical literature on the subject of graphs (\eg{}~\cite{New2018}), but there is also a significant body of more accessible, applied graph theory literature(see~\cite{FKL2019,Spo2018,ZCZetal2018,VMMR2017,FZB2016} for neuroscience-related reviews).
This applied literature relates the mathematical graph theory concepts to specific real world features of networks; however, it is important to remember that these real world meanings may not map one-to-one to the biology behind neuronal graphs, even from as closely a related field as fMRI studies.

In fact, making links between graph theory analysis and real-world biological meaning requires considerable understanding of both the mathematics, experiment and neuroscience.

There are two ways the community can address this problem:
\begin{enumerate}
    \item by working closely with graph theorists on projects to develop modified-algorithms that probe specific hypotheses and/or utilise \apriori{} biological knowledge to reveal new information, and
    \item by embedding graph theory and network science experts into groups developing and using calcium imaging techniques.
\end{enumerate}
Both of these approaches create an ongoing dialogue that ensures the appropriate approaches are used and that no underlying assumptions are broken.

Additionally, many of these metrics are best used in a comparative fashion with other real experiments or with \insilico{} controls, i.e. computationally created networks lacking true, information processing organisation.

\section*{Graph Models of Neurons}\label{sec:models}

Exploiting graph theory to analyse neuronal graphs enables quantitative comparison between different sample groups, \eg{} drug \vs{} no drug, by comparing metrics between graphs.
Probabilistic modelling of \emph{random graphs} also enables the comparison of real world neuronal graphs to to an \insilico{} control.
\Insilico{} controls allow scientists to compare neuronal graphs with random graphs that have similar properties, e.g. edge, degree distribution, etc., but lack any controlled organisation.
Such comparisons can be used to a) confirm that properties of neuronal graphs are statistically significant; b) provide a baseline from which different experiments can be compared and c) be used to guide the formation on new computational models that lead to a better understanding of neural mechanisms of computation, \eg{}~\cite{SBM2019}.

Comparisons between the topological structure of random graphs and real graphs has been used in the study of many complex networks across diverse disciplines.
In this Section, we will introduce three well established random graph models, which all display different topological structures and thus have different uses and limitations.

\subsection*{Random Networks --- The Erd\"{o}s-R\'{e}nyi Model}\label{subsec:models:random}

\begin{figure}
  \begin{minipage}[b][][b]{0.32\linewidth}
    \centering%
    \textbf{Erd\"{o}s-R\'{e}nyi}
  \end{minipage}
  \begin{minipage}[b][][b]{0.32\linewidth}
    \centering%
    \textbf{Watts-Strogatz}
  \end{minipage}
  \begin{minipage}[b][][b]{0.32\linewidth}
    \centering%
    \textbf{Barab\'{a}si-Albert}
  \end{minipage}%

  \vspace{5pt}%

  \setlength{\algomargin}{-0.1em}
  \begin{subfigure}[t][14em][t]{0.32\linewidth}
    \centering%
    \begin{algorithm}[H]
      \SetKwProg{ForEach}{foreach}{}{end}
      \SetKwFunction{KwRand}{rand}
      \SetKwProg{Generate}{generate}{ then}{}
      \SetAlgoLined%
      \DontPrintSemicolon%
      \Generate{empty graph}{}
      add \(N_V\) nodes\;
      \ForEach{pair of nodes}{
        \If{\KwRand{} > p}{add edge}
      }
    \end{algorithm}
    \caption{}\label{fig:models:er:flow}
  \end{subfigure}
  \begin{subfigure}[t][14em][t]{0.32\linewidth}
    \centering%
    \begin{algorithm}[H]
      \SetKwProg{Generate}{generate}{ then}{}
      \SetKwProg{ForEach}{foreach}{}{end}
      \SetKwFunction{KwRand}{rand}
      \SetAlgoLined%
      \DontPrintSemicolon%
      \Generate{ring lattice}{}
      \ForEach{node}{
        \ForEach{existing edge}{
          \If{\KwRand{} > p}{rewire edge}
        }
      }
    \end{algorithm}
    \caption{}\label{fig:models:ws:flow}
  \end{subfigure}
  \begin{subfigure}[t][14em][t]{0.32\linewidth}
    \centering%
    \begin{algorithm}[H]
      \SetKwProg{Generate}{generate}{ then}{}
      \SetKwProg{ForEach}{foreach}{}{end}
      \SetKwProg{While}{while}{}{end}
      \SetKwFunction{KwRand}{rand}
      \SetKwFunction{KwNum}{number}
      \SetAlgoLined%
      \DontPrintSemicolon%
      \Generate{empty graph}{}
      add m nodes\;
      \While{\KwNum{nodes} < \(N_V\)}{
        add node\;
        \ForEach{other node}{
          \If{\KwRand{} < k/K}{add edge}
        }
      }
    \end{algorithm}
    \caption{}\label{fig:models:ba:flow}
  \end{subfigure}%

  \begin{subfigure}[b][][b]{0.32\linewidth}
    \centering%
    \begin{tikzpicture}
	\GraphInit[vstyle=Simple]
	\SetGraphUnit{1}
	\tikzset{VertexStyle/.append style = {minimum size=5pt, fill=NavyBlue!0}}
	\Vertex[x=-0.41313271117781436,y=-0.4510381493894061]{V0}
	\AddVertexColor{NavyBlue!92.3076923076923}{V0}
	\Vertex[x=-0.23877701018196973,y=-1.91068033338261]{V1}
	\AddVertexColor{NavyBlue!38.46153846153847}{V1}
	\Vertex[x=-1.8346684718440658,y=0.7383870394827495]{V2}
	\AddVertexColor{NavyBlue!30.76923076923077}{V2}
	\Vertex[x=0.10107231831894511,y=1.4045223462223968]{V3}
	\AddVertexColor{NavyBlue!53.84615384615385}{V3}
	\Vertex[x=-1.2042726510979356,y=-0.2389357682976435]{V4}
	\AddVertexColor{NavyBlue!61.53846153846154}{V4}
	\Vertex[x=0.520787649045311,y=1.2029924441606479]{V5}
	\AddVertexColor{NavyBlue!53.84615384615385}{V5}
	\Vertex[x=0.14179179958554672,y=-0.6070449002507997]{V6}
	\AddVertexColor{NavyBlue!53.84615384615385}{V6}
	\Vertex[x=1.2003069164062294,y=0.11475474999976104]{V7}
	\AddVertexColor{NavyBlue!46.15384615384615}{V7}
	\Vertex[x=0.04503994981179593,y=0.19671370352279788]{V8}
	\AddVertexColor{NavyBlue!61.53846153846154}{V8}
	\Vertex[x=1.024201309007867,y=0.931875399548059]{V9}
	\AddVertexColor{NavyBlue!53.84615384615385}{V9}
	\Vertex[x=0.818718042618883,y=0.37027628250414657]{V10}
	\AddVertexColor{NavyBlue!61.53846153846154}{V10}
	\Vertex[x=0.9797632300800869,y=-1.7074010277063003]{V11}
	\AddVertexColor{NavyBlue!46.15384615384615}{V11}
	\Vertex[x=-1.2782558680145593,y=0.275530431318715]{V12}
	\AddVertexColor{NavyBlue!38.46153846153847}{V12}
	\Vertex[x=-0.34784411066371407,y=0.7716157682551951]{V13}
	\AddVertexColor{NavyBlue!61.53846153846154}{V13}
	\Vertex[x=-0.5249608170495419,y=-0.1398363855795123]{V14}
	\AddVertexColor{NavyBlue!53.84615384615385}{V14}
	\Vertex[x=1.534261655253622,y=-0.22987581730596676]{V15}
	\AddVertexColor{NavyBlue!38.46153846153847}{V15}
	\Vertex[x=-0.5475395442259275,y=1.1941976245096704]{V16}
	\AddVertexColor{NavyBlue!46.15384615384615}{V16}
	\Vertex[x=0.35708324938733504,y=-1.7616605359556863]{V17}
	\AddVertexColor{NavyBlue!46.15384615384615}{V17}
	\Vertex[x=-0.44345375405940524,y=1.6812749599103884]{V18}
	\AddVertexColor{NavyBlue!30.76923076923077}{V18}
	\Vertex[x=0.10161724806194805,y=-0.26123072770638545]{V19}
	\AddVertexColor{NavyBlue!53.84615384615385}{V19}
	\Vertex[x=0.3964055347869598,y=1.8534030685261969]{V20}
	\AddVertexColor{NavyBlue!38.46153846153847}{V20}
	\Vertex[x=-1.354010193073578,y=-0.9537488476011352]{V21}
	\AddVertexColor{NavyBlue!38.46153846153847}{V21}
	\Vertex[x=2.0,y=-1.0568724636341227]{V22}
	\AddVertexColor{NavyBlue!30.76923076923077}{V22}
	\Vertex[x=1.0493969998945243,y=-0.5447469071167886]{V23}
	\AddVertexColor{NavyBlue!61.53846153846154}{V23}
	\Vertex[x=-0.3981635788911453,y=-1.4729026945410897]{V24}
	\AddVertexColor{NavyBlue!61.53846153846154}{V24}
	\Vertex[x=0.9948902016946711,y=1.4975219092370784]{V25}
	\AddVertexColor{NavyBlue!38.46153846153847}{V25}
	\Vertex[x=-0.992084234207855,y=1.1059704031289173]{V26}
	\AddVertexColor{NavyBlue!53.84615384615385}{V26}
	\Vertex[x=-1.5823655077792553,y=-1.5027083477246717]{V27}
	\AddVertexColor{NavyBlue!23.076923076923077}{V27}
	\Vertex[x=-0.5537780383902561,y=0.2361763006202564]{V28}
	\AddVertexColor{NavyBlue!69.23076923076923}{V28}
	\Vertex[x=0.4479703867032984,y=-0.7365295247548672]{V29}
	\AddVertexColor{NavyBlue!100.0}{V29}
	\Edge[local=true,lw=1.0pt,,style={opacity=0.5}](V0)(V2)
	\Edge[local=true,lw=1.0pt,,style={opacity=0.5}](V0)(V4)
	\Edge[local=true,lw=1.0pt,,style={opacity=0.5}](V0)(V8)
	\Edge[local=true,lw=1.0pt,,style={opacity=0.5}](V0)(V10)
	\Edge[local=true,lw=1.0pt,,style={opacity=0.5}](V0)(V11)
	\Edge[local=true,lw=1.0pt,,style={opacity=0.5}](V0)(V13)
	\Edge[local=true,lw=1.0pt,,style={opacity=0.5}](V0)(V14)
	\Edge[local=true,lw=1.0pt,,style={opacity=0.5}](V0)(V17)
	\Edge[local=true,lw=1.0pt,,style={opacity=0.5}](V0)(V20)
	\Edge[local=true,lw=1.0pt,,style={opacity=0.5}](V0)(V24)
	\Edge[local=true,lw=1.0pt,,style={opacity=0.5}](V0)(V27)
	\Edge[local=true,lw=1.0pt,,style={opacity=0.5}](V0)(V28)
	\Edge[local=true,lw=1.0pt,,style={opacity=0.5}](V1)(V14)
	\Edge[local=true,lw=1.0pt,,style={opacity=0.5}](V1)(V17)
	\Edge[local=true,lw=1.0pt,,style={opacity=0.5}](V1)(V19)
	\Edge[local=true,lw=1.0pt,,style={opacity=0.5}](V1)(V24)
	\Edge[local=true,lw=1.0pt,,style={opacity=0.5}](V1)(V29)
	\Edge[local=true,lw=1.0pt,,style={opacity=0.5}](V2)(V4)
	\Edge[local=true,lw=1.0pt,,style={opacity=0.5}](V2)(V13)
	\Edge[local=true,lw=1.0pt,,style={opacity=0.5}](V2)(V16)
	\Edge[local=true,lw=1.0pt,,style={opacity=0.5}](V3)(V5)
	\Edge[local=true,lw=1.0pt,,style={opacity=0.5}](V3)(V8)
	\Edge[local=true,lw=1.0pt,,style={opacity=0.5}](V3)(V9)
	\Edge[local=true,lw=1.0pt,,style={opacity=0.5}](V3)(V10)
	\Edge[local=true,lw=1.0pt,,style={opacity=0.5}](V3)(V14)
	\Edge[local=true,lw=1.0pt,,style={opacity=0.5}](V3)(V20)
	\Edge[local=true,lw=1.0pt,,style={opacity=0.5}](V3)(V26)
	\Edge[local=true,lw=1.0pt,,style={opacity=0.5}](V4)(V8)
	\Edge[local=true,lw=1.0pt,,style={opacity=0.5}](V4)(V14)
	\Edge[local=true,lw=1.0pt,,style={opacity=0.5}](V4)(V19)
	\Edge[local=true,lw=1.0pt,,style={opacity=0.5}](V4)(V21)
	\Edge[local=true,lw=1.0pt,,style={opacity=0.5}](V4)(V26)
	\Edge[local=true,lw=1.0pt,,style={opacity=0.5}](V4)(V29)
	\Edge[local=true,lw=1.0pt,,style={opacity=0.5}](V5)(V7)
	\Edge[local=true,lw=1.0pt,,style={opacity=0.5}](V5)(V10)
	\Edge[local=true,lw=1.0pt,,style={opacity=0.5}](V5)(V16)
	\Edge[local=true,lw=1.0pt,,style={opacity=0.5}](V5)(V19)
	\Edge[local=true,lw=1.0pt,,style={opacity=0.5}](V5)(V25)
	\Edge[local=true,lw=1.0pt,,style={opacity=0.5}](V5)(V26)
	\Edge[local=true,lw=1.0pt,,style={opacity=0.5}](V6)(V7)
	\Edge[local=true,lw=1.0pt,,style={opacity=0.5}](V6)(V8)
	\Edge[local=true,lw=1.0pt,,style={opacity=0.5}](V6)(V13)
	\Edge[local=true,lw=1.0pt,,style={opacity=0.5}](V6)(V16)
	\Edge[local=true,lw=1.0pt,,style={opacity=0.5}](V6)(V17)
	\Edge[local=true,lw=1.0pt,,style={opacity=0.5}](V6)(V23)
	\Edge[local=true,lw=1.0pt,,style={opacity=0.5}](V6)(V24)
	\Edge[local=true,lw=1.0pt,,style={opacity=0.5}](V7)(V9)
	\Edge[local=true,lw=1.0pt,,style={opacity=0.5}](V7)(V14)
	\Edge[local=true,lw=1.0pt,,style={opacity=0.5}](V7)(V23)
	\Edge[local=true,lw=1.0pt,,style={opacity=0.5}](V7)(V29)
	\Edge[local=true,lw=1.0pt,,style={opacity=0.5}](V8)(V10)
	\Edge[local=true,lw=1.0pt,,style={opacity=0.5}](V8)(V12)
	\Edge[local=true,lw=1.0pt,,style={opacity=0.5}](V8)(V23)
	\Edge[local=true,lw=1.0pt,,style={opacity=0.5}](V8)(V29)
	\Edge[local=true,lw=1.0pt,,style={opacity=0.5}](V9)(V13)
	\Edge[local=true,lw=1.0pt,,style={opacity=0.5}](V9)(V15)
	\Edge[local=true,lw=1.0pt,,style={opacity=0.5}](V9)(V16)
	\Edge[local=true,lw=1.0pt,,style={opacity=0.5}](V9)(V20)
	\Edge[local=true,lw=1.0pt,,style={opacity=0.5}](V9)(V29)
	\Edge[local=true,lw=1.0pt,,style={opacity=0.5}](V10)(V18)
	\Edge[local=true,lw=1.0pt,,style={opacity=0.5}](V10)(V22)
	\Edge[local=true,lw=1.0pt,,style={opacity=0.5}](V10)(V23)
	\Edge[local=true,lw=1.0pt,,style={opacity=0.5}](V10)(V29)
	\Edge[local=true,lw=1.0pt,,style={opacity=0.5}](V11)(V17)
	\Edge[local=true,lw=1.0pt,,style={opacity=0.5}](V11)(V22)
	\Edge[local=true,lw=1.0pt,,style={opacity=0.5}](V11)(V23)
	\Edge[local=true,lw=1.0pt,,style={opacity=0.5}](V11)(V24)
	\Edge[local=true,lw=1.0pt,,style={opacity=0.5}](V11)(V29)
	\Edge[local=true,lw=1.0pt,,style={opacity=0.5}](V12)(V18)
	\Edge[local=true,lw=1.0pt,,style={opacity=0.5}](V12)(V21)
	\Edge[local=true,lw=1.0pt,,style={opacity=0.5}](V12)(V28)
	\Edge[local=true,lw=1.0pt,,style={opacity=0.5}](V12)(V29)
	\Edge[local=true,lw=1.0pt,,style={opacity=0.5}](V13)(V19)
	\Edge[local=true,lw=1.0pt,,style={opacity=0.5}](V13)(V20)
	\Edge[local=true,lw=1.0pt,,style={opacity=0.5}](V13)(V26)
	\Edge[local=true,lw=1.0pt,,style={opacity=0.5}](V13)(V28)
	\Edge[local=true,lw=1.0pt,,style={opacity=0.5}](V14)(V18)
	\Edge[local=true,lw=1.0pt,,style={opacity=0.5}](V14)(V24)
	\Edge[local=true,lw=1.0pt,,style={opacity=0.5}](V15)(V19)
	\Edge[local=true,lw=1.0pt,,style={opacity=0.5}](V15)(V22)
	\Edge[local=true,lw=1.0pt,,style={opacity=0.5}](V15)(V28)
	\Edge[local=true,lw=1.0pt,,style={opacity=0.5}](V15)(V29)
	\Edge[local=true,lw=1.0pt,,style={opacity=0.5}](V16)(V26)
	\Edge[local=true,lw=1.0pt,,style={opacity=0.5}](V16)(V28)
	\Edge[local=true,lw=1.0pt,,style={opacity=0.5}](V17)(V19)
	\Edge[local=true,lw=1.0pt,,style={opacity=0.5}](V17)(V29)
	\Edge[local=true,lw=1.0pt,,style={opacity=0.5}](V18)(V25)
	\Edge[local=true,lw=1.0pt,,style={opacity=0.5}](V19)(V26)
	\Edge[local=true,lw=1.0pt,,style={opacity=0.5}](V20)(V25)
	\Edge[local=true,lw=1.0pt,,style={opacity=0.5}](V21)(V24)
	\Edge[local=true,lw=1.0pt,,style={opacity=0.5}](V21)(V28)
	\Edge[local=true,lw=1.0pt,,style={opacity=0.5}](V21)(V29)
	\Edge[local=true,lw=1.0pt,,style={opacity=0.5}](V22)(V29)
	\Edge[local=true,lw=1.0pt,,style={opacity=0.5}](V23)(V24)
	\Edge[local=true,lw=1.0pt,,style={opacity=0.5}](V23)(V25)
	\Edge[local=true,lw=1.0pt,,style={opacity=0.5}](V23)(V29)
	\Edge[local=true,lw=1.0pt,,style={opacity=0.5}](V24)(V27)
	\Edge[local=true,lw=1.0pt,,style={opacity=0.5}](V25)(V28)
	\Edge[local=true,lw=1.0pt,,style={opacity=0.5}](V26)(V28)
	\Edge[local=true,lw=1.0pt,,style={opacity=0.5}](V27)(V28)
\end{tikzpicture}%
    \caption{}\label{fig:models:er:graph}
  \end{subfigure}
  \begin{subfigure}[b][][b]{0.32\linewidth}
    \centering%
    \begin{tikzpicture}
	\GraphInit[vstyle=Simple]
	\SetGraphUnit{1}
	\tikzset{VertexStyle/.append style = {minimum size=5pt, fill=NavyBlue!0}}
	\Vertex[x=-1.394465743867995,y=1.1397990368642072]{V0}
	\AddVertexColor{NavyBlue!23.076923076923077}{V0}
	\Vertex[x=-0.6162410713027733,y=0.8285814157776678]{V1}
	\AddVertexColor{NavyBlue!30.76923076923077}{V1}
	\Vertex[x=-0.6842036626434987,y=0.4560125436596294]{V2}
	\AddVertexColor{NavyBlue!46.15384615384615}{V2}
	\Vertex[x=-0.1931294211844977,y=0.781680604189744]{V3}
	\AddVertexColor{NavyBlue!23.076923076923077}{V3}
	\Vertex[x=0.9812938236999567,y=2.0]{V4}
	\AddVertexColor{NavyBlue!15.384615384615385}{V4}
	\Vertex[x=0.828866965526081,y=0.9080492858299946]{V5}
	\AddVertexColor{NavyBlue!30.76923076923077}{V5}
	\Vertex[x=0.21725540489572417,y=1.6093640376990723]{V6}
	\AddVertexColor{NavyBlue!38.46153846153847}{V6}
	\Vertex[x=0.07551297590323726,y=0.8608180069006788]{V7}
	\AddVertexColor{NavyBlue!30.76923076923077}{V7}
	\Vertex[x=0.3065823888139807,y=1.1325732072389034]{V8}
	\AddVertexColor{NavyBlue!38.46153846153847}{V8}
	\Vertex[x=1.1010719696357303,y=1.3221276954363952]{V9}
	\AddVertexColor{NavyBlue!23.076923076923077}{V9}
	\Vertex[x=1.1828985665269423,y=0.23495038356653977]{V10}
	\AddVertexColor{NavyBlue!38.46153846153847}{V10}
	\Vertex[x=1.2246736737778208,y=-0.9385649844432087]{V11}
	\AddVertexColor{NavyBlue!30.76923076923077}{V11}
	\Vertex[x=0.7010728112674773,y=-0.6249599035266249]{V12}
	\AddVertexColor{NavyBlue!30.76923076923077}{V12}
	\Vertex[x=1.0869291710863853,y=-0.35378466440197137]{V13}
	\AddVertexColor{NavyBlue!30.76923076923077}{V13}
	\Vertex[x=0.20256644680257338,y=-0.23555660425955996]{V14}
	\AddVertexColor{NavyBlue!38.46153846153847}{V14}
	\Vertex[x=1.4471071813435084,y=-0.464085275233617]{V15}
	\AddVertexColor{NavyBlue!23.076923076923077}{V15}
	\Vertex[x=-0.4312553040323025,y=1.768424120933062]{V16}
	\AddVertexColor{NavyBlue!23.076923076923077}{V16}
	\Vertex[x=-0.5616946608131317,y=-1.950003868936019]{V17}
	\AddVertexColor{NavyBlue!23.076923076923077}{V17}
	\Vertex[x=-0.36191451341790215,y=-0.8796211065931696]{V18}
	\AddVertexColor{NavyBlue!38.46153846153847}{V18}
	\Vertex[x=0.40019929729569975,y=-1.8752776021418298]{V19}
	\AddVertexColor{NavyBlue!23.076923076923077}{V19}
	\Vertex[x=0.17671645642453912,y=-0.8877879489511303]{V20}
	\AddVertexColor{NavyBlue!46.15384615384615}{V20}
	\Vertex[x=-0.2255245188285649,y=-1.252815027715334]{V21}
	\AddVertexColor{NavyBlue!30.76923076923077}{V21}
	\Vertex[x=-0.3571776988048837,y=-0.3564964006195588]{V22}
	\AddVertexColor{NavyBlue!30.76923076923077}{V22}
	\Vertex[x=-0.6482827164767768,y=-0.20065431537939954]{V23}
	\AddVertexColor{NavyBlue!53.84615384615385}{V23}
	\Vertex[x=-1.4664985471799203,y=-0.13303031806363302]{V24}
	\AddVertexColor{NavyBlue!23.076923076923077}{V24}
	\Vertex[x=-0.9411316645396561,y=-0.8001391656798197]{V25}
	\AddVertexColor{NavyBlue!38.46153846153847}{V25}
	\Vertex[x=-1.323660427730402,y=-1.4720229353431524]{V26}
	\AddVertexColor{NavyBlue!23.076923076923077}{V26}
	\Vertex[x=1.2169724372377781,y=-1.2097703400712738]{V27}
	\AddVertexColor{NavyBlue!23.076923076923077}{V27}
	\Vertex[x=-1.0536240972707658,y=-0.4245608433338917]{V28}
	\AddVertexColor{NavyBlue!23.076923076923077}{V28}
	\Vertex[x=-0.8909155221443619,y=1.0167509665972958]{V29}
	\AddVertexColor{NavyBlue!30.76923076923077}{V29}
	\Edge[local=true,lw=1.0pt,,style={opacity=0.5}](V0)(V1)
	\Edge[local=true,lw=1.0pt,,style={opacity=0.5}](V0)(V29)
	\Edge[local=true,lw=1.0pt,,style={opacity=0.5}](V0)(V2)
	\Edge[local=true,lw=1.0pt,,style={opacity=0.5}](V1)(V3)
	\Edge[local=true,lw=1.0pt,,style={opacity=0.5}](V1)(V8)
	\Edge[local=true,lw=1.0pt,,style={opacity=0.5}](V1)(V22)
	\Edge[local=true,lw=1.0pt,,style={opacity=0.5}](V2)(V23)
	\Edge[local=true,lw=1.0pt,,style={opacity=0.5}](V2)(V6)
	\Edge[local=true,lw=1.0pt,,style={opacity=0.5}](V2)(V18)
	\Edge[local=true,lw=1.0pt,,style={opacity=0.5}](V2)(V7)
	\Edge[local=true,lw=1.0pt,,style={opacity=0.5}](V2)(V24)
	\Edge[local=true,lw=1.0pt,,style={opacity=0.5}](V3)(V5)
	\Edge[local=true,lw=1.0pt,,style={opacity=0.5}](V3)(V23)
	\Edge[local=true,lw=1.0pt,,style={opacity=0.5}](V4)(V5)
	\Edge[local=true,lw=1.0pt,,style={opacity=0.5}](V4)(V6)
	\Edge[local=true,lw=1.0pt,,style={opacity=0.5}](V5)(V7)
	\Edge[local=true,lw=1.0pt,,style={opacity=0.5}](V5)(V13)
	\Edge[local=true,lw=1.0pt,,style={opacity=0.5}](V6)(V8)
	\Edge[local=true,lw=1.0pt,,style={opacity=0.5}](V6)(V16)
	\Edge[local=true,lw=1.0pt,,style={opacity=0.5}](V6)(V9)
	\Edge[local=true,lw=1.0pt,,style={opacity=0.5}](V7)(V16)
	\Edge[local=true,lw=1.0pt,,style={opacity=0.5}](V7)(V14)
	\Edge[local=true,lw=1.0pt,,style={opacity=0.5}](V8)(V9)
	\Edge[local=true,lw=1.0pt,,style={opacity=0.5}](V8)(V10)
	\Edge[local=true,lw=1.0pt,,style={opacity=0.5}](V8)(V29)
	\Edge[local=true,lw=1.0pt,,style={opacity=0.5}](V9)(V10)
	\Edge[local=true,lw=1.0pt,,style={opacity=0.5}](V10)(V11)
	\Edge[local=true,lw=1.0pt,,style={opacity=0.5}](V10)(V12)
	\Edge[local=true,lw=1.0pt,,style={opacity=0.5}](V10)(V15)
	\Edge[local=true,lw=1.0pt,,style={opacity=0.5}](V11)(V12)
	\Edge[local=true,lw=1.0pt,,style={opacity=0.5}](V11)(V13)
	\Edge[local=true,lw=1.0pt,,style={opacity=0.5}](V11)(V19)
	\Edge[local=true,lw=1.0pt,,style={opacity=0.5}](V12)(V14)
	\Edge[local=true,lw=1.0pt,,style={opacity=0.5}](V12)(V18)
	\Edge[local=true,lw=1.0pt,,style={opacity=0.5}](V13)(V27)
	\Edge[local=true,lw=1.0pt,,style={opacity=0.5}](V13)(V20)
	\Edge[local=true,lw=1.0pt,,style={opacity=0.5}](V14)(V25)
	\Edge[local=true,lw=1.0pt,,style={opacity=0.5}](V14)(V23)
	\Edge[local=true,lw=1.0pt,,style={opacity=0.5}](V14)(V15)
	\Edge[local=true,lw=1.0pt,,style={opacity=0.5}](V15)(V27)
	\Edge[local=true,lw=1.0pt,,style={opacity=0.5}](V16)(V29)
	\Edge[local=true,lw=1.0pt,,style={opacity=0.5}](V17)(V18)
	\Edge[local=true,lw=1.0pt,,style={opacity=0.5}](V17)(V19)
	\Edge[local=true,lw=1.0pt,,style={opacity=0.5}](V17)(V26)
	\Edge[local=true,lw=1.0pt,,style={opacity=0.5}](V18)(V20)
	\Edge[local=true,lw=1.0pt,,style={opacity=0.5}](V18)(V25)
	\Edge[local=true,lw=1.0pt,,style={opacity=0.5}](V19)(V21)
	\Edge[local=true,lw=1.0pt,,style={opacity=0.5}](V20)(V21)
	\Edge[local=true,lw=1.0pt,,style={opacity=0.5}](V20)(V22)
	\Edge[local=true,lw=1.0pt,,style={opacity=0.5}](V20)(V27)
	\Edge[local=true,lw=1.0pt,,style={opacity=0.5}](V20)(V28)
	\Edge[local=true,lw=1.0pt,,style={opacity=0.5}](V21)(V22)
	\Edge[local=true,lw=1.0pt,,style={opacity=0.5}](V21)(V23)
	\Edge[local=true,lw=1.0pt,,style={opacity=0.5}](V22)(V23)
	\Edge[local=true,lw=1.0pt,,style={opacity=0.5}](V23)(V24)
	\Edge[local=true,lw=1.0pt,,style={opacity=0.5}](V23)(V25)
	\Edge[local=true,lw=1.0pt,,style={opacity=0.5}](V24)(V25)
	\Edge[local=true,lw=1.0pt,,style={opacity=0.5}](V25)(V26)
	\Edge[local=true,lw=1.0pt,,style={opacity=0.5}](V26)(V28)
	\Edge[local=true,lw=1.0pt,,style={opacity=0.5}](V28)(V29)
\end{tikzpicture}%
    \caption{}\label{fig:models:ws:graph}
  \end{subfigure}
  \begin{subfigure}[b][][b]{0.32\linewidth}
    \centering%
    \begin{tikzpicture}
	\GraphInit[vstyle=Simple]
	\SetGraphUnit{1}
	\tikzset{VertexStyle/.append style = {minimum size=5pt, fill=NavyBlue!0}}
	\Vertex[x=-1.1587141098834761,y=-0.3046303541136033]{V0}
	\AddVertexColor{NavyBlue!76.92307692307693}{V0}
	\Vertex[x=0.320477698644445,y=0.18081562753869426]{V1}
	\AddVertexColor{NavyBlue!100.0}{V1}
	\Vertex[x=0.11049848754595781,y=-0.5309599444316346]{V2}
	\AddVertexColor{NavyBlue!92.3076923076923}{V2}
	\Vertex[x=-0.8491374868701248,y=-1.3262799775395437]{V3}
	\AddVertexColor{NavyBlue!38.46153846153847}{V3}
	\Vertex[x=0.10175817729669823,y=0.5652546165170483]{V4}
	\AddVertexColor{NavyBlue!84.61538461538461}{V4}
	\Vertex[x=1.0815372398385819,y=0.932611111632104]{V5}
	\AddVertexColor{NavyBlue!46.15384615384615}{V5}
	\Vertex[x=0.6856553631943471,y=-0.08362479048051077]{V6}
	\AddVertexColor{NavyBlue!15.384615384615385}{V6}
	\Vertex[x=-0.5730918992622535,y=-0.38885676340689646]{V7}
	\AddVertexColor{NavyBlue!15.384615384615385}{V7}
	\Vertex[x=-1.0504510990668896,y=-1.085473638038763]{V8}
	\AddVertexColor{NavyBlue!30.76923076923077}{V8}
	\Vertex[x=-0.22545236829848977,y=-1.7588667449650397]{V9}
	\AddVertexColor{NavyBlue!15.384615384615385}{V9}
	\Vertex[x=-0.27134622631252914,y=-0.9622805078678627]{V10}
	\AddVertexColor{NavyBlue!15.384615384615385}{V10}
	\Vertex[x=1.3345902750679435,y=-0.007526288874424569]{V11}
	\AddVertexColor{NavyBlue!15.384615384615385}{V11}
	\Vertex[x=1.0416842677581821,y=1.4741025886644583]{V12}
	\AddVertexColor{NavyBlue!15.384615384615385}{V12}
	\Vertex[x=-0.8901693008348605,y=0.9378403651914254]{V13}
	\AddVertexColor{NavyBlue!23.076923076923077}{V13}
	\Vertex[x=-2.0,y=-0.8982614269751975]{V14}
	\AddVertexColor{NavyBlue!15.384615384615385}{V14}
	\Vertex[x=0.9457267817666326,y=0.5609956632783553]{V15}
	\AddVertexColor{NavyBlue!15.384615384615385}{V15}
	\Vertex[x=0.554405528131027,y=-0.6879917179235084]{V16}
	\AddVertexColor{NavyBlue!23.076923076923077}{V16}
	\Vertex[x=-0.5826594790202685,y=0.09908486632844195]{V17}
	\AddVertexColor{NavyBlue!15.384615384615385}{V17}
	\Vertex[x=-1.3115815786806777,y=0.6171013169099948]{V18}
	\AddVertexColor{NavyBlue!15.384615384615385}{V18}
	\Vertex[x=0.3809031416856074,y=1.3753944761920578]{V19}
	\AddVertexColor{NavyBlue!23.076923076923077}{V19}
	\Vertex[x=-0.9873128199563344,y=0.14566998768914663]{V20}
	\AddVertexColor{NavyBlue!15.384615384615385}{V20}
	\Vertex[x=0.9349748760823727,y=-1.338441594795808]{V21}
	\AddVertexColor{NavyBlue!23.076923076923077}{V21}
	\Vertex[x=1.6025597691793259,y=0.6752664841679974]{V22}
	\AddVertexColor{NavyBlue!15.384615384615385}{V22}
	\Vertex[x=-0.940580110172874,y=0.5828712683880889]{V23}
	\AddVertexColor{NavyBlue!15.384615384615385}{V23}
	\Vertex[x=0.9377446750079259,y=1.8322334791288246]{V24}
	\AddVertexColor{NavyBlue!23.076923076923077}{V24}
	\Vertex[x=1.3776138695688076,y=-0.8203944266874298]{V25}
	\AddVertexColor{NavyBlue!15.384615384615385}{V25}
	\Vertex[x=-0.37553457766224735,y=1.3173792353234444]{V26}
	\AddVertexColor{NavyBlue!15.384615384615385}{V26}
	\Vertex[x=0.5662817226256797,y=0.8744298033422683]{V27}
	\AddVertexColor{NavyBlue!15.384615384615385}{V27}
	\Vertex[x=0.9458293614239685,y=-0.6173869182012289]{V28}
	\AddVertexColor{NavyBlue!15.384615384615385}{V28}
	\Vertex[x=-1.706210178796474,y=-1.3600757959909044]{V29}
	\AddVertexColor{NavyBlue!15.384615384615385}{V29}
	\Edge[local=true,lw=1.0pt,,style={opacity=0.5}](V0)(V2)
	\Edge[local=true,lw=1.0pt,,style={opacity=0.5}](V0)(V3)
	\Edge[local=true,lw=1.0pt,,style={opacity=0.5}](V0)(V7)
	\Edge[local=true,lw=1.0pt,,style={opacity=0.5}](V0)(V13)
	\Edge[local=true,lw=1.0pt,,style={opacity=0.5}](V0)(V14)
	\Edge[local=true,lw=1.0pt,,style={opacity=0.5}](V0)(V17)
	\Edge[local=true,lw=1.0pt,,style={opacity=0.5}](V0)(V18)
	\Edge[local=true,lw=1.0pt,,style={opacity=0.5}](V0)(V20)
	\Edge[local=true,lw=1.0pt,,style={opacity=0.5}](V0)(V23)
	\Edge[local=true,lw=1.0pt,,style={opacity=0.5}](V0)(V29)
	\Edge[local=true,lw=1.0pt,,style={opacity=0.5}](V1)(V2)
	\Edge[local=true,lw=1.0pt,,style={opacity=0.5}](V1)(V4)
	\Edge[local=true,lw=1.0pt,,style={opacity=0.5}](V1)(V5)
	\Edge[local=true,lw=1.0pt,,style={opacity=0.5}](V1)(V7)
	\Edge[local=true,lw=1.0pt,,style={opacity=0.5}](V1)(V10)
	\Edge[local=true,lw=1.0pt,,style={opacity=0.5}](V1)(V15)
	\Edge[local=true,lw=1.0pt,,style={opacity=0.5}](V1)(V17)
	\Edge[local=true,lw=1.0pt,,style={opacity=0.5}](V1)(V19)
	\Edge[local=true,lw=1.0pt,,style={opacity=0.5}](V1)(V20)
	\Edge[local=true,lw=1.0pt,,style={opacity=0.5}](V1)(V22)
	\Edge[local=true,lw=1.0pt,,style={opacity=0.5}](V1)(V25)
	\Edge[local=true,lw=1.0pt,,style={opacity=0.5}](V1)(V26)
	\Edge[local=true,lw=1.0pt,,style={opacity=0.5}](V1)(V28)
	\Edge[local=true,lw=1.0pt,,style={opacity=0.5}](V2)(V3)
	\Edge[local=true,lw=1.0pt,,style={opacity=0.5}](V2)(V4)
	\Edge[local=true,lw=1.0pt,,style={opacity=0.5}](V2)(V6)
	\Edge[local=true,lw=1.0pt,,style={opacity=0.5}](V2)(V8)
	\Edge[local=true,lw=1.0pt,,style={opacity=0.5}](V2)(V9)
	\Edge[local=true,lw=1.0pt,,style={opacity=0.5}](V2)(V11)
	\Edge[local=true,lw=1.0pt,,style={opacity=0.5}](V2)(V16)
	\Edge[local=true,lw=1.0pt,,style={opacity=0.5}](V2)(V21)
	\Edge[local=true,lw=1.0pt,,style={opacity=0.5}](V2)(V27)
	\Edge[local=true,lw=1.0pt,,style={opacity=0.5}](V2)(V28)
	\Edge[local=true,lw=1.0pt,,style={opacity=0.5}](V3)(V8)
	\Edge[local=true,lw=1.0pt,,style={opacity=0.5}](V3)(V9)
	\Edge[local=true,lw=1.0pt,,style={opacity=0.5}](V3)(V29)
	\Edge[local=true,lw=1.0pt,,style={opacity=0.5}](V4)(V5)
	\Edge[local=true,lw=1.0pt,,style={opacity=0.5}](V4)(V6)
	\Edge[local=true,lw=1.0pt,,style={opacity=0.5}](V4)(V12)
	\Edge[local=true,lw=1.0pt,,style={opacity=0.5}](V4)(V13)
	\Edge[local=true,lw=1.0pt,,style={opacity=0.5}](V4)(V15)
	\Edge[local=true,lw=1.0pt,,style={opacity=0.5}](V4)(V16)
	\Edge[local=true,lw=1.0pt,,style={opacity=0.5}](V4)(V18)
	\Edge[local=true,lw=1.0pt,,style={opacity=0.5}](V4)(V19)
	\Edge[local=true,lw=1.0pt,,style={opacity=0.5}](V4)(V23)
	\Edge[local=true,lw=1.0pt,,style={opacity=0.5}](V5)(V11)
	\Edge[local=true,lw=1.0pt,,style={opacity=0.5}](V5)(V12)
	\Edge[local=true,lw=1.0pt,,style={opacity=0.5}](V5)(V22)
	\Edge[local=true,lw=1.0pt,,style={opacity=0.5}](V5)(V24)
	\Edge[local=true,lw=1.0pt,,style={opacity=0.5}](V8)(V10)
	\Edge[local=true,lw=1.0pt,,style={opacity=0.5}](V8)(V14)
	\Edge[local=true,lw=1.0pt,,style={opacity=0.5}](V13)(V26)
	\Edge[local=true,lw=1.0pt,,style={opacity=0.5}](V16)(V21)
	\Edge[local=true,lw=1.0pt,,style={opacity=0.5}](V19)(V24)
	\Edge[local=true,lw=1.0pt,,style={opacity=0.5}](V21)(V25)
	\Edge[local=true,lw=1.0pt,,style={opacity=0.5}](V24)(V27)
\end{tikzpicture}%
    \caption{}\label{fig:models:ba:graph}
  \end{subfigure}%

  \begin{subfigure}[b][][b]{0.32\linewidth}
    \centering%
    \includegraphics[width=\linewidth]{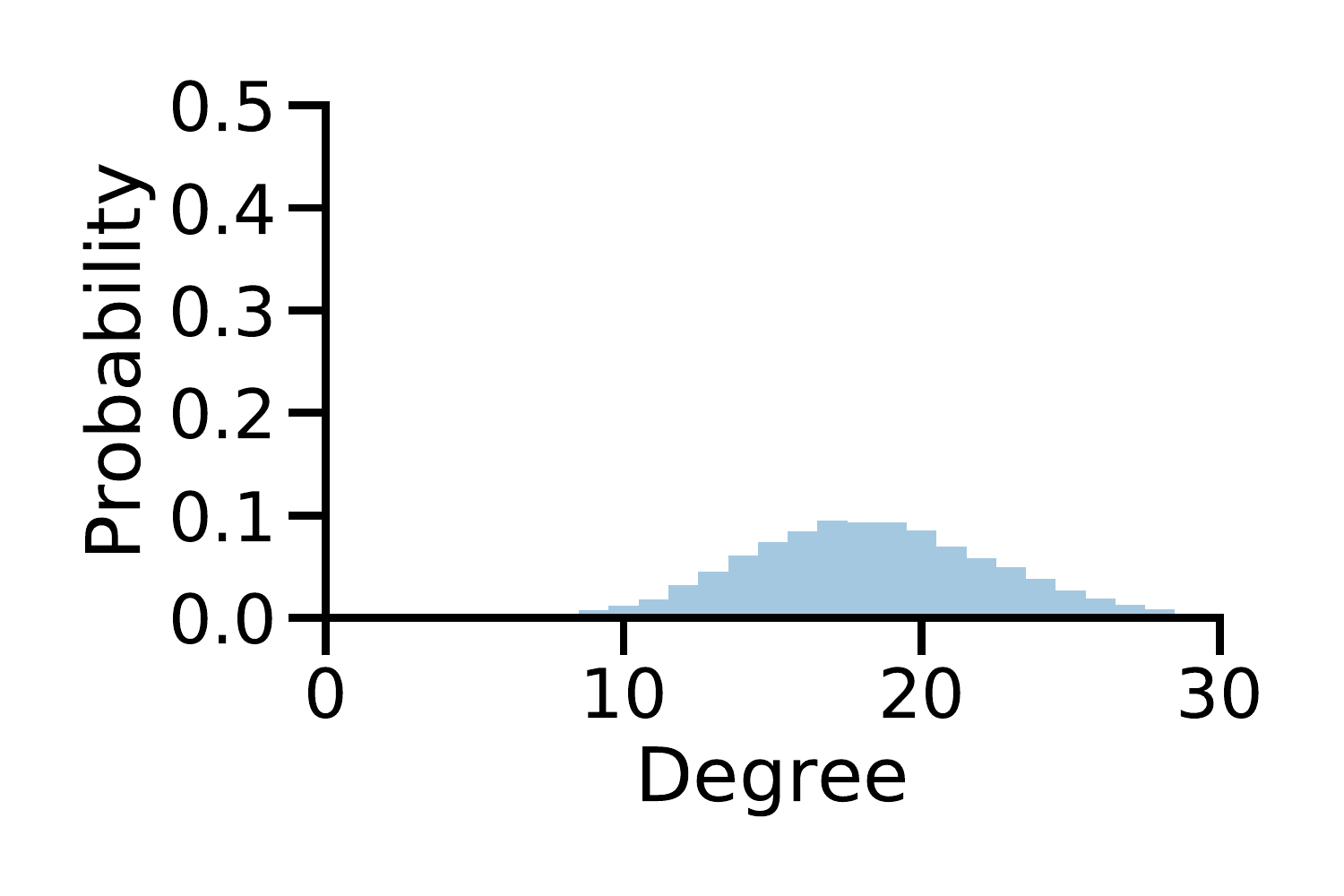}
    \caption{}\label{fig:models:er:plot}
  \end{subfigure}
  \begin{subfigure}[b][][b]{0.32\linewidth}
    \centering%
    \includegraphics[width=\linewidth]{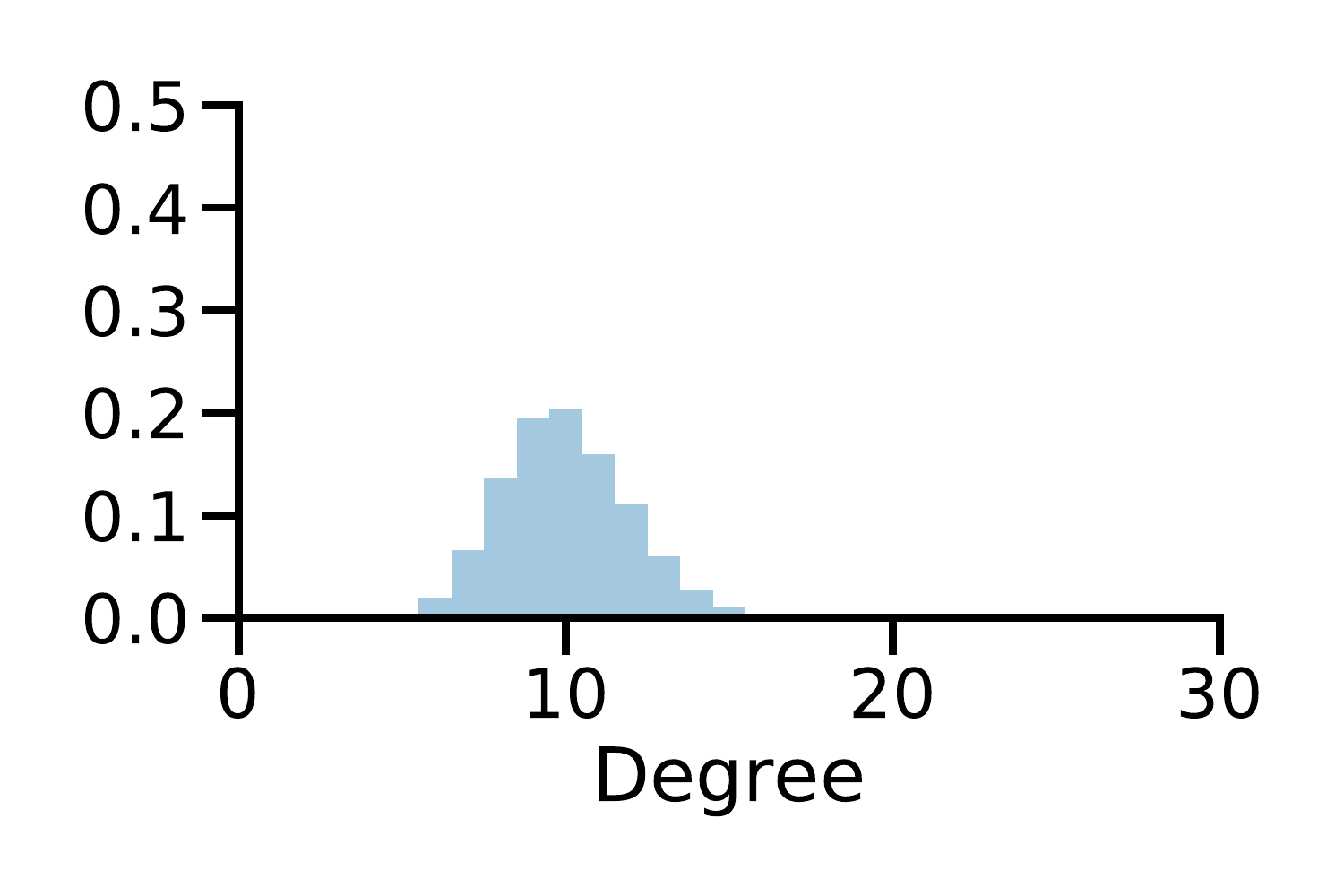}
    \caption{}\label{fig:models:ws:plot}
  \end{subfigure}
  \begin{subfigure}[b][][b]{0.32\linewidth}
    \centering%
    \includegraphics[width=\linewidth]{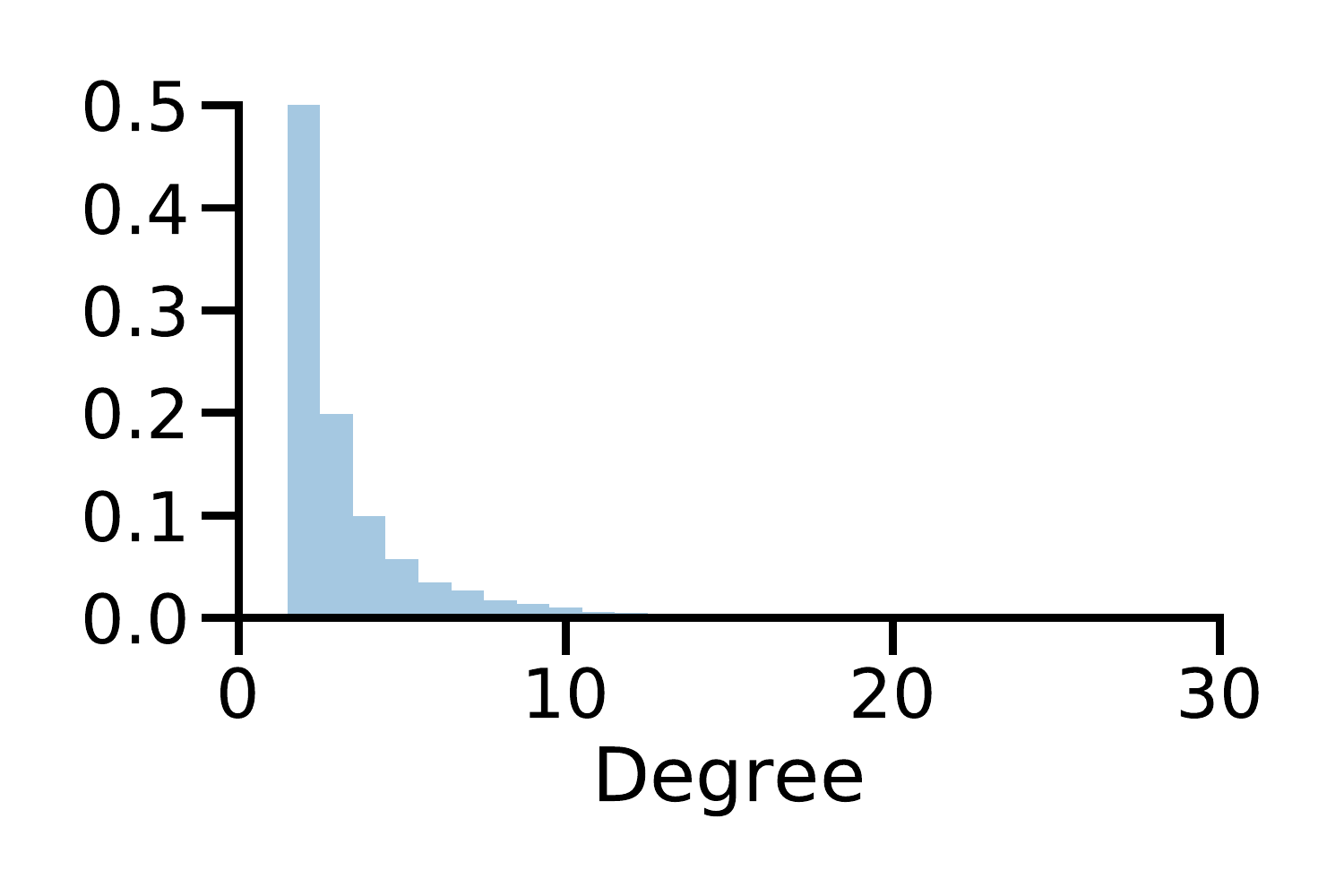}
    \caption{}\label{fig:models:ba:plot}
  \end{subfigure}%
  \caption{\textbf{Random graphs, which can be used as models or controls, can be generated in different ways giving the graphs different properties.} (\textbf{A--C}) Pseudocode showing the processes used to create Erd\"{o}s-R\'{e}nyi (\textbf{A}), Watts-Strogatz (\textbf{B}) and Barab\'{a}si-Albert (\textbf{C}) model graphs. (\textbf{D--F}) Example graphs with \(N_V = 30\) showing clearly different organisations for different generation models. (\textbf{G--I}) Probability distributions of node degree over graphs generated with \(N_V = 10,000\) showing lower average degree and increased tails in both the Watts-Strogatz (\textbf{H}) and Barab\'{a}si-Albert (\textbf{I}) models.}\label{fig:graph-models}
\end{figure}

In the Erd\"{o}s-R\'{e}nyi (ER) model~\cite{ER1959, Gil1959}, a graph \(G\) with \(N_V\) nodes is constructed by connecting pairs of nodes, \eg{} \( \{ u, v \} \), randomly with probability \(p\).
The creation of every edge, \(E_{u,v}\), is independent from all other edges, \ie{} each edge is randomly added regardless of other edges that have or have not been created (\cref{fig:models:er:flow}).

The ER model generate homogenous, random graphs (\cref{fig:models:er:graph}); however, they assume that edges are independent, which is not true in biological systems.
Unlike neuronal graphs, ER graphs do not display local clustering of nodes nor do they show small-world properties seen in many real-world and biological systems, as shown in the zebrafish~\cite{APMetal2017,BHGB2019}.

In fMRI data, ER graphs and functional brain networks have been compared using graph metrics and modelled temporal dynamics.
~\cite{BHV2016} showed that functional brain networks from fMRI show different topological properties to density-matched ER graphs.
Further, they showed that modelling BOLD activity on both real and ER graphs showed dissimilar results, indicating the importance of network organisation on dynamic signalling.
Thus ER graphs are good random graphs but don't accurately represent many graphs found in the real world \cite{LKF2005}.

\subsection*{Small-World Networks --- The Watts-Strogatz Model}\label{subsec:models:small-world}

The Watts-Strogatz (WS) model~\cite{WS1998} was designed to generate random graphs whilst accounting for, and replicating, features seen in real-world systems.
Specifically, the WS model was designed to maintain the low average shortest path lengths of the ER model whilst increasing local clustering coefficient (compared to the ER model).

\begin{figure}
  \centering
      \begin{tikzpicture}
    \begin{scope}
        \GraphInit[vstyle=Simple]
        \SetGraphUnit{1}
        \tikzset{VertexStyle/.append style = {minimum size=8pt, fill=NavyBlue!0}}
        \Vertices{circle}{V1,V2,V3,V4,V5,V6}
        \AddVertexColor{NavyBlue}{V1}
        \Edge[local=true,lw=2.0pt,color=NavyBlue](V1)(V2)
        \Edge[local=true,lw=2.0pt,color=NavyBlue](V1)(V3)

        \Edge[local=true,lw=2.0pt](V2)(V3)
        \Edge[local=true,lw=2.0pt](V2)(V4)

        \Edge[local=true,lw=2.0pt](V3)(V4)
        \Edge[local=true,lw=2.0pt](V3)(V5)

        \Edge[local=true,lw=2.0pt](V4)(V5)
        \Edge[local=true,lw=2.0pt](V4)(V6)

        \Edge[local=true,lw=2.0pt](V5)(V6)
        \Edge[local=true,lw=2.0pt,color=NavyBlue](V5)(V1)

        \Edge[local=true,lw=2.0pt,color=NavyBlue](V6)(V1)
        \Edge[local=true,lw=2.0pt](V6)(V2)

      \end{scope}%
\end{tikzpicture}%
  \caption{A 4-Regular Ring Lattice on a 6 node graph. The blue edges connected to node 1 show why this graph is 4-regular graph: All nodes have exactly 4 edges connecting them to their 4 closet neighbours}\label{fig:ring_lattice}
\end{figure}
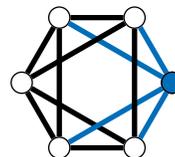

In the WS model, a ring lattice graph \(G\) (An example of such a graph is highlighted in Figure \ref{fig:ring_lattice}) with \(N_V\) nodes, where each node is connected to its \(k\) nearest neighbour nodes only, is generated.
For each node, each of it's existing edges is rewired (randomly) with probability \(\beta \)

WS graphs are heterogeneous and vary in randomness, based on parameters, usually with high modularity.
WS graphs show `small-world' properties, where nodes that are not neighbours are connected by a short path (\cf{} Six Degrees of Kevin Bacon).

Many real networks show small-world topologies.
In neuroscience, small-world networks are an attractive model as their local clustering allows for highly localised processing while their low average shortest path lengths also allows for distributed processing~\cite{BB2016}.
This balance of local and distributed information processing allows small-world networks to be highly efficient for minimal wiring cost.

\subsubsection*{Measuring Small-Worldness}\label{subsubsec:models:small-world:small-worldness}

It is possible to measure how `small-world' a graph is by comparing the graph clustering and path lengths of that graph to that of an equivalent but randomly generated graph.
Most simply one can calculate the small-coefficient \(\sigma \) (also known as the small-world quotient~\cite{DYB2003,US2005}),
\begin{equation}\label{eq:sigma}
  \sigma = \frac{~\frac{C_G}{C_R}~}{~\frac{l_G}{l_R}~} ,
\end{equation}
where \(C\) and \(l\) are the clustering coefficient and average shortest path length of graph \(G\) and random graph \(R\)~\cite{KW2001}.
Graphs where \(\sigma>1\) are small-world.
However, the small-coefficient is influenced by the size of the graph in question~\cite{Nea2017}.
Because of this the small-coefficient is not good for comparing different graphs.

Alternatively, one can use the small-world measures \(\omega \) or \(\omega'\), the latter of which provides a measure between \(0\) and \(1\).
A \(0\) small-world measure indicates a graph is as `unsmall-worldly' as can be (given the graph size, degree, \etc{}), whereas \(1\) indicates a graph is as small-worldly as possible.
The small world measure relates the properties of graph \(G\) to an equivalent lattice graph \(\ell \) (completely ordered and non-random) and random graph \(R\)~\cite{TJHetal2011},
\begin{equation}\label{eq:omega}
  \omega = \frac{l_R}{l_G} - \frac{C_G}{C_\ell}, \quad \omega' = 1 - \left|\omega\right| .
\end{equation}
The small-world measure is good for comparing two graphs with similar properties; however, the range of results depends on other constraints on the graph, \eg{} density, degree distribution and more, and so two graphs that differ on these constraints may both have \(\omega'=1\) but may not be equally close to a theoretical ideal of a small world network~\cite{Nea2017}.

An alternative measure was proposed by~\cite{Nea2015} --- the small world index (\(SWI\)), defined as
\begin{equation}\label{eq:swi}
  SWI = \frac{l_G - l_\ell}{l_R - l_\ell} \times \frac{C_G - C_R}{C_\ell - C_R} .
\end{equation}
Like \(\omega \), \(SWI\) ranges between \(0\) and \(1\) where \(1\) indicates a graph with theoretically ideal small-world characteristics given other constraints on the graph (as with \(\omega\) above).
Of these three metrics the \(SWI\) most closely matches the WS definition of a small-world graph.

A similar metric, the small world propensity, was proposed by~\cite{MBB2016}, defined as
\begin{equation}\label{eq:swp}
    \phi = 1 - \sqrt{\frac{\Delta_C^2 + \Delta_l^2}{2}} ,
\end{equation}
where \(\Delta_C = \frac{C_\ell - C_G}{C_\ell - C_R}\) and \(\Delta_l = \frac{l_G - l_R}{l_\ell - l_R}\). Both \(\Delta_C\) and \(\Delta_\ell\) are bounded between \num{0} and \num{1}.

Like the \(SWI\), \(\phi\) ranges between \(0\) and \(1\) where \(1\) indicates a graph with high small-world characteristics.
The small world propensity was designed to provide an unbiased assessment of small world structure in brain networks regardless of graph density.
The small world propensity can be extended for weighted graphs and both the weighted and unweighted variants can be used to generate model graphs.

Due to the computational constraints of large, whole-brain networks, a simplified version of small-worldliness was measured in~\cite{BHGB2019}, where the authors showed a significant difference in small-worldliness in brains of wild-type and `depressed' zebrafish (\emph{gr\textsuperscript{s357}} mutants) exposed to different antidepressant drugs.

It's worth noting that~\cite{MBB2016} showed that the weighted whole-\textit{C. elegans} neuronal graph did not show a high small-world propensity.
The authors argue that this could be as the whole-animal neuronal graph does not just represent the head and that the organism is evolutionary simple compared to other model organisms.
The authors recommend stringent examination of small-world feature across scale, physiology and evolutionary scales.

\subsection*{Scale-Free Networks --- The Barab\'{a}si-Albert Model}\label{subsec:models:scale-free}

As the average shortest path length becomes smaller these small-world networks can begin to show scale-free properties.
In a scale-free network, the degree distribution follows a power law, \ie{} \(P(k) \sim k^{-\gamma}\).
Scale-free networks have a small number of very connected nodes (large degree) and a large number of less connected nodes (small degree), creating a long right-tailed degree distribution~\cite{CSN2009}.

The Barab\'{a}si-Albert (BA) model~\cite{AB2002} was designed to generate random graphs with a scale-free (power-law) degree distribution.
Specifically, the BA model incorporates a preferential attachment mechanism to generate graphs that share properties with many real-world networks, possibly including networks of neurons~\cite{SMU2014,APMetal2017}.

In the BA model, a small graph of \(m\) nodes is created.
Nodes are then added one at a time until the total number of nodes \(N_V\) is reached.
After each new node is added, an edge is created between the new node and an existing node \(i\) with probability \(k_i/{\Sigma k}\), \ie{} new edges are preferentially created with existing nodes with high degree (\(k_i\)).

BA graphs are heterogeneous, with a small number of nodes having a relatively high number of connections (high degree), whilst the majority of nodes have a low degree.
This process naturally results in a high clustering coefficients and hub-like nodes.

In~\cite{APMetal2017}, the authors suggest that the network topology in the zebrafish optic tectum is scale-free and show that the degree distribution fits a power law.
In their research, neural assemblies in the optic tectum stay scale-free throughout development and despite other changes in network topology due to dark-rearing.

However, there is a growing body of evidence that this type of graph is less common in real world systems~\cite{BC2019}.
Indeed, changing the threshold used to generate a neuronal graph can influence the degree distribution significantly, as shown in zebrafish whole-brain imaging~\cite{BHGB2019} and neuronal cell cultures~\cite{SMU2014}.
We would like to suggest caution in the interpretation of graphs as scale-free and suggest researchers follow the rigorous protocols suggested by~\cite{BC2019}.

\subsubsection*{A Critique of Scale-Free Graphs}\label{subsubsec:models:scale-free:critique}

Before moving onto the identification of scale-free networks, it is worth considering the cautionary tales present in recent literature.
Since the original observations made by Barab\'{a}si that many empirical networks demonstrate scale-free degree distributions~\cite{BA1999}, numerous other researchers have also measured the property in everything from citation to road networks~\cite{KSCM2006}. 

However, there has been a growing body of work demonstrating that the scale-free property might not be as prevalent in the real world as first imagined.
For example, work by ~\cite{BC2019} has shown that in nearly one thousand empirical network datasets, constructed from data across a broad range of scientific disciplines, only a tiny fraction are actually scale-free (when using a strict definition of the property).
The paper calls into question the universality of the scale-free topologies, with biological networks being one of the network classes to display the least number of scale-free examples.
These ideas are not new, and earlier work has also argued against the prevalence of scale-free networks in the real world~\cite{JH2003}.
Conversely it has also been argued that the concept of scale-free networks can still be useful, even in light of these new discoveries~\cite{H2019}.

As such, the question of how to recognise data that obeys a power law is a tricky one.
\Cref{box:scalefree} summarises the statistically rigorous process recommended by~\cite{CSN2009} for the case of a graph with degree distribution that may or may not follow a power law.

\begin{sidebox}[label={box:scalefree}]{Identifying Scale-Free Graphs.}
  The question of how to recognise data that obeys a power law is a tricky one.
  A statistically rigorous set of statistical techniques was proposed by~\cite{CSN2009}, in which they also showed a number of real world data that had been miss-ascribed as scale-free.
  \begin{enumerate}
    \item First, we define our model degree distribution as,
      \begin{equation}\label{eq:power-law}
        P(k) \sim \frac{k^{-\alpha}}{\zeta(\alpha,k_{min})}, \quad\text{where}\quad \zeta(\alpha,k_{min})=\sum_{n=0}^\infty {(n + k_{min})}^{-\alpha}.
      \end{equation}
      We then estimate the power law model parameters \(\alpha \) and \(k_{min}\).
      \(k_{min}\) is the minimum number degree for which the power law applies and \(\alpha \) is the scaling parameter.
      Data that follows an ideal power law across the whole domain will have \(k_{min}=1\).
    \item In order to determine the lower bound \(k_{min}\) we compare the real data with the model above \(k_{min}\) using the Kolmogorov-Smirnov (KV) statistic.
      To calculate exactly the scaling parameter \(\alpha \) we can use discrete maximum likelihood estimators.
      However, for ease of evaluation we can also approximate \(\alpha \) where a discrete distribution, like degree distribution, is treated as a continuous distribution rounded to the nearest integer.
    
    \item Next we calculate the goodness-of-fit between the data and the model.
      This is achieved by generating the a large number of synthetic datasets that follow the determined power law model.
      We then measure the KV statistic between the model and our synthetic data and compare this to the KV statistic between the model and our real data.
      The \(p\)-value is then the fraction of times that \(KV_{synthetic}\) is greater than \(KV_{real}\).
      A \(p\)-value above \(0.1\) indicates that the power law is plausible, whilst a value below \(p\) below \(0.1\) indicates that the model should be rejected.
    
    \item Finally, we compare the data to other models, \eg{} exponential, Poisson or Yule distributions through likelihood ratio tests; techniques such as cross-validation or Bayesian approaches could also be used.
  \end{enumerate}
\end{sidebox}

\subsection*{Machine Learning Generated Networks}\label{subsec:models:ml-gen}

The recent advances in machine learning on graphs, specifically the family of Graph Neural Network (GNN) models~\cite{HYL2017}, has resulted in new methods for generating random graphs based on a set of input training graphs.
Whilst there have, thus far, been limited possibilities for applications in biology, we briefly review some of the more prominent approaches and encourage readers to investigate further.

A family of neural-based generative models entitled auto-encoders have been adapted to generate random graphs. 
Auto-encoders are a type of artificial neural network which learn a low dimensional representation of input data, which is used to then reconstruct the original data \cite{HS2006}.
They are frequently combined with techniques from variational inference to create Variational Auto-Encoders (VAE), which improves the reconstruction and interpretability of the model \cite{KW2013}.
Recent graph generation approaches, such as GraphVAE \cite{SK2018}, Junction Tree VAE \cite{JBJ2018} and others \cite{MCX2018, ZJCGC2019}, all produce a model which is trained on a given set of input graphs and can then generate random synthetic examples which have a similar topological structure to the input set.

Using an alternative approach and underlying model, GraphRNN \cite{YYRHL2018} exploits Recurrent Neural Networks (RNN), a type of model designed for sequential tasks \cite{HS1997}, for graph generation.
GraphRNN again learns to produce graphs matching the topologies of a set of input graphs, but this time by learning to decompose the graph generation problem into a sequence of node and edge placements.g

\section*{Graph Analysis Tools}

There are a wide range of tools for visualising and quantifying graphs.
In this section we briefly introduce a few open-source projects that may be useful for researchers who wish to explore graph theory in their work.

First, it's important to understand the possible challenges that one might face dealing with large graphs of complex data.
Consider a zebrafish whole-brain imaging experiment, one dataset might contain up to \num{100000} neurons, so a graph of the whole brain would have \num{100000} nodes.
If a graph was then created with weighted edges between all nodes (a total of \num{10000000000} edges) then the adjacency matrix alone, \ie{} without any additional information such as position in the brain, would take \SI{80}{\giga\byte}.
Graph visualisation or analysis tools may require the loading of all or most of this data in the computer memory (RAM), thus a powerful computer is required.

This problem may be overcome by considering a subset of nodes or of edges, and many tasks can be programmed to account for memory or processing concerns.

\subsection*{Graph Visualisation and Analysis Software}

Graph visualisation (or graph drawing) has been a large area of research in it's own right and a wide range of software and programming packages exist for calculating graph layouts and visualising massive graph data.

Two notable open-source tools are Cytoscape~\cite{SMOetal2003} and Gephi~\cite{BHJ2009}.
Cytoscape was designed for visualising molecular networks in 'omics studies.
Cytoscape has a wide range of visualisation tools including fast rendering and easy live navigating of large graphs.
Plug-ins are available to enable graph filtering, identification of clusters and other tasks.

Gephi is designed for a more general audience and has been used in research from social networks through digital humanities and in journalism.
Like Cytoscape, Gephi is able to carry out real-time visualisation and exploration of large graphs, and has built-in tools to calculate a variety of metrics.
Gephi is also supported by community-built plug-ins.

\subsection*{Programming with Graphs}

In many cases it may be beneficial to work with neuronal graphs within existing pipelines and programming environments.
There are many tools available for different programming languages that provide graph visualisation and analysis functions.
Several notable Python modules exist that are open-source, are well-documented and easy to use.

NetworkX~\cite{HSS2008} is perhaps the most well known and has a more gentle learning curve than the others.
Many of the metrics we've described in this paper are built-in to NetworkX, along with a variety input/output options, visualisation tools and random graph generators.
Importantly, NetworkX is well-documented and still in active development.
However, NetworkX is designed, primarily, for small graphs and the many-edged, massive nature of some neuronal graphs may prove a challenge.

Other well documented Python modules for the analysis of massive graphs are graph-tool~\cite{Pei2017} and NetworKit~\cite{SSM2015}.
Aware of the computational challenges of processing large graphs, graph-tool makes use of the increased performance available by using algorithms implemented in C++ however, they keep the usability of a Python front-end.
Similarly, NetworKit uses a combination of C++, parallelisation and heuristic algorithms to deal with computational expensive problems in a fast and parallel manner.
As such, both can provide an alternative (or complement) to NetworkX for analysing and visualising large graphs in Python.

However, as graphs continue to grow in both complexity and size, there is an increasing need to scale graph computation from one machine to many.
Parallel computation packages, such as Apache Spark~\cite{Z2012}, have greatly reduced many of the complexities traditionally associated with parallel programming, whilst also offering a python interface for ease of use.
Apache Spark offers a selection of graph specific frameworks, such as GraphX~\cite{G2014} and GraphFrames~\cite{AAL2016}, which include implicit algorithms to extract many of the metrics discussed in the metrics section in parallel across a compute cluster.
Unfortunately, for any metric not included by default, one must be coded in accordance with Spark's parallel programming model, still a non-trivial task.

Additionally, exploiting GPU technologies may bring massive changes in the speed of processing and the maximum graph sizes that can be handled.
Again, these currently require custom algorithms; however, recently a new package has been released that begin to address this problem - Rapids cuGraph~\cite{Tea2018}.

\section*{Concluding Remarks}

We hope that the reader can see the power of using neuronal graphs to explore network organisation as opposed to considering only population statistics.
In particular we hope that the reader appreciates that the metrics presented in this paper are only the tip of the iceberg and that any interested researcher should identify and build collaborations with graph theorists and network scientists to ensure that the right metric is used to answer the question of interest (see~\cite{KBBetal2015,VB2007} for best practice).
Further, we hope the reader recognises that, with the use of graph models, it is possible to compare against \insilico{} controls for null hypothesis testing in order to ensure robust and statistically rigorous conclusions are drawn.
We greatly look forward to the coming years and seeing more and novel applications of graph theory in calcium imaging, and can see neuronal graphs becoming a powerful tool that helps to both answer biological questions and pose new biological hypotheses.
We strongly predict that graph theory analysis on calcium imaging will lead to important new insights that allow neuroscientists to understand and model computational networks.

\bibliography{main}

\end{document}